\documentclass{article}

\usepackage{PRIMEarxiv}

\usepackage[utf8]{inputenc}
\usepackage[T1]{fontenc}
\usepackage{hyperref}
\usepackage{url}
\usepackage{booktabs}
\usepackage{amsmath}
\usepackage{amssymb}
\usepackage{amsfonts}
\usepackage{nicefrac}
\usepackage{microtype}
\usepackage{graphicx}
\usepackage{multirow}
\usepackage{adjustbox}
\usepackage{enumitem}
\usepackage{stfloats}
\usepackage{placeins}
\usepackage{fancyhdr}
\usepackage{authblk}
\usepackage{algorithm}
\usepackage{algorithmic}

\graphicspath{{media/}}

\DeclareUnicodeCharacter{00A0}{ }

\newcommand{\mstd}[2]{#1 $\pm$ #2}

\fancypagestyle{empty}{
  \fancyhf{}
  \lfoot{This paper is accepted by IEEE Sensors Journal.} 
  
}
\title{STAMBRIDGE: Spectral-Temporal Amplitude-aware Mid-Feature Bridge for EEG Visual Decoding%
\thanks{\textit{\underline{Citation}}: Jiahe Meng, Weiming Zeng, Yueyang Li, Bo Chai, Hongjie Yan, Zhiguo Zhang, Wai Ting Siok, Nizhuan Wang. STAMBRIDGE: Spectral-Temporal Amplitude-aware Mid-Feature Bridge for EEG Visual Decoding [J], IEEE Sensors Journal, 2026.}}

\author[1]{Jiahe Meng}
\author[1]{Weiming Zeng*}
\author[1,2]{Yueyang Li}
\author[2]{Bo Chai}
\author[3]{Hongjie Yan}
\author[4]{Zhiguo Zhang}
\author[2]{Wai Ting Siok}
\author[2]{Nizhuan Wang*}

\affil[1]{Lab of Digital Image and Intelligent Computation, Shanghai Maritime University, Shanghai 201306, China}
\affil[2]{Department of Language Science and Technology, The Hong Kong Polytechnic University, Hung Hom, Kowloon, Hong Kong SAR, China}
\affil[3]{Department of Neurology, Affiliated Lianyungang Hospital of Xuzhou Medical University, Lianyungang 222002, China}
\affil[4]{Institute of Computing and Intelligence, Harbin Institute of Technology Shenzhen, Shenzhen 518000, China}

\date{}

\begin{document}
\maketitle

\begin{abstract}
Decoding visual stimuli using electroencephalography (EEG) remains especially challenging due to the significant modality gap between low-SNR neural signals and highly structured vision--language spaces. Unlike auditory signals, which share temporal information with EEG, the high-dimensional spatial nature of visual data makes direct cross-modal alignment inherently more unstable than that of other modalities. To address this, we propose STAMBRIDGE, a versatile two-stage framework that sequentially tackles feature conditioning and cross-modal alignment. First, we introduce a Spectral-Temporal Amplitude-aware Modulation (STAM) to extract well-conditioned EEG representations. By replacing hard frequency masking with amplitude-derived soft channel weighting and multi-scale temporal convolutions, STAM explicitly preserves frequency-aware transients while reducing the risk of time-domain ringing artifacts. Building upon these robust neural features, we further introduce a model-agnostic Mid-Feature Semantic Bridge (MFSB) that constructs a regularized intermediate space through directed cross-modal interactions, enabling staged distillation and more stable semantic alignment. Experiments on the THINGS-EEG benchmark show competitive performance in 200-way zero-shot retrieval, with 34.50\% Top-1 and 65.95\% Top-5 accuracy. In addition, embeddings learned by STAMBRIDGE produce semantically coherent image reconstructions with a diffusion model, demonstrating robust EEG-to-vision semantic alignment. The code is available at: \url{https://github.com/thabeatmjh/STAMBRIDGE}.
\end{abstract}

\vspace{10pt}
\begin{center}
  \textbf{Graphical Abstract}\\
  \vspace{6pt}
  \includegraphics[width=0.85\linewidth]{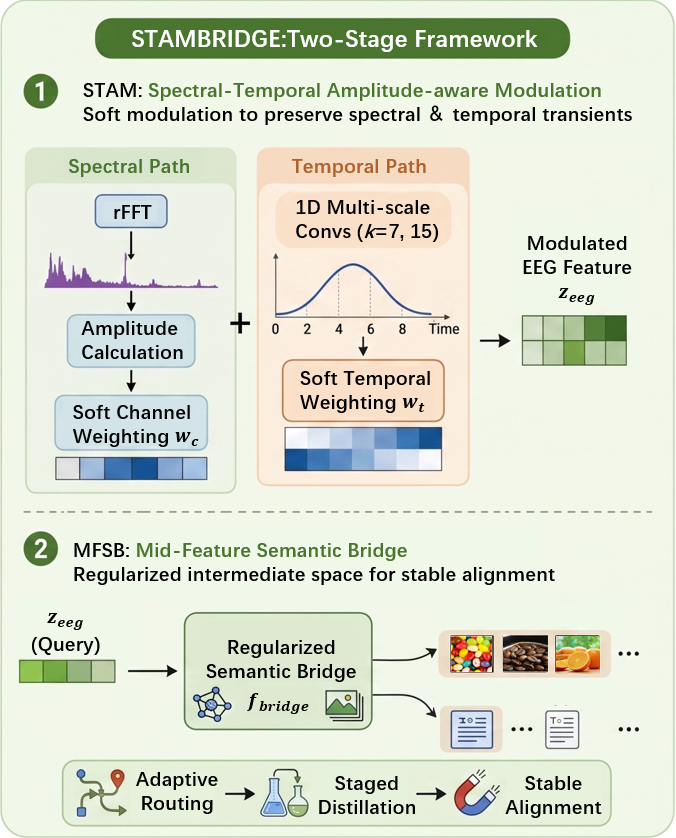}
\end{center}
\vspace{10pt}

\keywords{EEG-based visual decoding \and multimodal contrastive representation learning \and spectral--temporal modulation \and mid-feature distillation}

\section{Introduction}
\label{sec:intro}

Decoding visual content from non-invasive brain recordings is a fundamental problem in both cognitive neuroscience and next-generation brain--computer interfaces (BCIs)\cite{li2025tale}. Among available neuroimaging modalities, electroencephalography (EEG) is particularly attractive due to its low cost, portability, and high temporal resolution, making it suitable for latency-sensitive applications \cite{willett2021high}. Nevertheless, EEG-based visual decoding remains highly challenging because EEG signals inherently exhibit low signal-to-noise ratio (SNR), coarse spatial resolution, and substantial inter-subject variability \cite{gibson2022eeg, xu2020crossdataset}. For visual decoding, these limitations are further compounded by the modality gap between unstructured brainwaves and highly organized visual-language latent spaces. Consequently, unlike auditory processing where EEG signals natively share temporal alignment with soundwaves, recovering high-level visual semantics from EEG requires bridging a massive dimensional and structural divide, making it considerably more difficult than from other neural measurements with higher SNR or finer spatial resolution.

A central prerequisite for EEG-based visual decoding is the extraction of neural representations that simultaneously preserve global temporal context and localized neurophysiological dynamics. Existing temporal modeling paradigms commonly rely on generic sequence architectures, such as the iTransformer \cite{liu2024itransformer}, which are effective at modeling long-range dependencies but may insufficiently capture localized frequency-aware transients. To incorporate spectral priors, recent studies such as Neural-MCRL \cite{li2025neural} introduce frequency-domain modulation mechanisms based on Fast Fourier Transforms (FFT). Although effective in enhancing spectral attention, these methods typically apply hard binary frequency masks before inverse transformations (IFFT). From a signal processing perspective, such abrupt spectral truncation inevitably introduces ringing artifacts in the time domain \cite{widmann2012filter, widmann2015digital}. The resulting temporal smearing can obscure subtle and short-lived neurophysiological transients that are critical for visual decoding. This issue becomes particularly severe in Rapid Serial Visual Presentation (RSVP) paradigms, where visual stimuli are presented at high frequencies and semantic information is encoded within precise stimulus-locked transient responses. Once these transients are distorted by filtering artifacts, the temporal discriminability between consecutive visual concepts is substantially weakened. To address this, we propose a Spectral-Temporal Amplitude-aware Modulation (STAM) module. Rather than acting as a conventional hard filter, STAM softly integrates spectral priors to extract well-conditioned representations while explicitly preserving the time-domain integrity of neural transients.

Beyond robust EEG representation learning, another major challenge lies in aligning neural embeddings with pretrained vision--language representations such as CLIP (Contrastive Language-Image Pre-training) \cite{radford2021learning}. As highlighted in recent comprehensive surveys \cite{li2025tbme}, although generative models have significantly advanced visual reconstruction quality, the dominant bottleneck remains the severe modality gap between EEG signals and semantic visual representations \cite{eslami2024mitigate}. Compared with the semantically structured vision--language embedding space, EEG signals—even after sophisticated encoding—remain only indirectly associated with high-level visual concepts. Existing approaches, including ATM \cite{li2024visual} and Neural-MCRL \cite{li2025neural}, commonly enforce direct one-step alignment between EEG embeddings and vision--language representations. However, this paradigm tightly couples several fundamentally different learning objectives within a single optimization stage, requiring the EEG encoder to simultaneously perform signal denoising, semantic abstraction, and cross-modal geometric alignment. As a consequence, the optimization process becomes dominated by the large gradient updates required for cross-modal matching, which can overwhelm the underlying neural representation learning. This optimization conflict makes it difficult to form a coherent latent structure and often leads to instability and semantic fragmentation, ultimately limiting retrieval and reconstruction performance. Unlike Neural-MCRL, which directly aligns EEG and visual representations within a single contrastive optimization stage, STAMBRIDGE explicitly decouples representation learning from semantic alignment through a staged optimization strategy.

To address aforementioned challenges, we proposed the STAMBRIDGE framework in this paper, whose key contributions are summarized as follows:
\begin{itemize}
\item We design an EEG encoding pipeline that combines a Subject-Specific Linear Layer ~\cite{wu2022transfer, li2019domain}, an iTransformer backbone, and the proposed Spectral-Temporal Amplitude-aware Modulation (STAM) module, which integrates multi-scale temporal convolutions and amplitude-derived soft channel weighting to capture frequency-aware transient dynamics while reducing filtering-induced temporal distortion.
\item We propose a Mid-Feature Semantic Bridge (MFSB), a model-agnostic alignment module that constructs a regularized intermediate semantic space for staged distillation, thereby alleviating optimization instability caused by the cross-modal gap.
\item We evaluate STAMBRIDGE on the THINGS-EEG benchmark, where it achieves competitive zero-shot retrieval performance. Furthermore, qualitative validation using an off-the-shelf latent diffusion model produces semantically coherent image reconstructions, demonstrating the improved alignment quality of the learned representations.
\end{itemize}

\section{Related Work}
\label{sec:related_work}

\subsection{EEG-based Visual Decoding}
Visual decoding traditionally relies on fMRI for its high spatial resolution \cite{scotti2024reconstructing, takagi2023high}. However, low temporal resolution and high operational costs limit its real-time BCI practicality. Consequently, EEG has emerged as a portable, millisecond-resolution alternative \cite{willett2021high}. Recent EEG decoding has shifted from coarse classification \cite{grootswagers2019representational} toward fine-grained retrieval and reconstruction. To model complex EEG dynamics, recent approaches increasingly use generic sequence backbones like Transformers and EEG-Conformer \cite{song2022eeg}. Although effective at global temporal modeling, they often overlook localized frequency-aware neurophysiological characteristics. Prior attempts to inject spectral priors typically rely on hard frequency masking, which may introduce time-domain ringing artifacts and distort short-lived neural signatures. This issue is particularly severe in Rapid Serial Visual Presentation (RSVP) paradigms \cite{gifford2022large}, where temporal discriminability is crucial, motivating a focus on more robust representation learning \cite{song2024decoding, li2025neural}.

\subsection{Multimodal Contrastive Learning \& Cross-Modal Alignment}
Vision--language models like CLIP \cite{radford2021learning} provide a robust shared semantic space for multimodal alignment. Recent zero-shot frameworks—including NICE \cite{song2024decoding}, BraVL \cite{du2023decoding}, ATM \cite{li2024visual}, and Neural-MCRL \cite{li2025neural}—project EEG embeddings toward pretrained semantic representations via contrastive learning, laying the foundation for modern EEG-to-vision decoding. Recently, this cross-modal alignment paradigm has even been extended beyond 2D natural images to 3D visual domains, such as latent EEG-vision alignment for 3D object reconstruction \cite{xiang2026latent3d}. 

To mitigate the remaining \textit{modality gap}, frameworks often rely on heavy external semantic augmentation, utilizing Retrieval-Augmented Generation (RAG) \cite{kim2025seeeeg} or Multimodal Large Language Models (MLLMs) \cite{xu2026mindsae}. While effective, these introduce substantial latency. Alternatively, internal alignment mechanisms like those in NeuroBridge \cite{zhang2026neurobridge} rely on coupled projectors that can propagate optimization pressure back to the fragile EEG backbone. Furthermore, while latent diffusion models \cite{rombach2022high, guo2025neuro3d, huang2025need} significantly advance visual reconstruction \cite{li2024visual}, their quality remains bottlenecked by upstream neural representations. In low-SNR paradigms like RSVP, models relying on coarse global conditioning struggle to preserve fine-grained semantics \cite{xiang2026regional}, highlighting the need for well-conditioned EEG embeddings that explicitly preserve spectral-temporal transients.

\begin{figure*}[!t]
  \centering 
  \adjustbox{max width=\textwidth, max height=0.42\textheight}{%
    \includegraphics{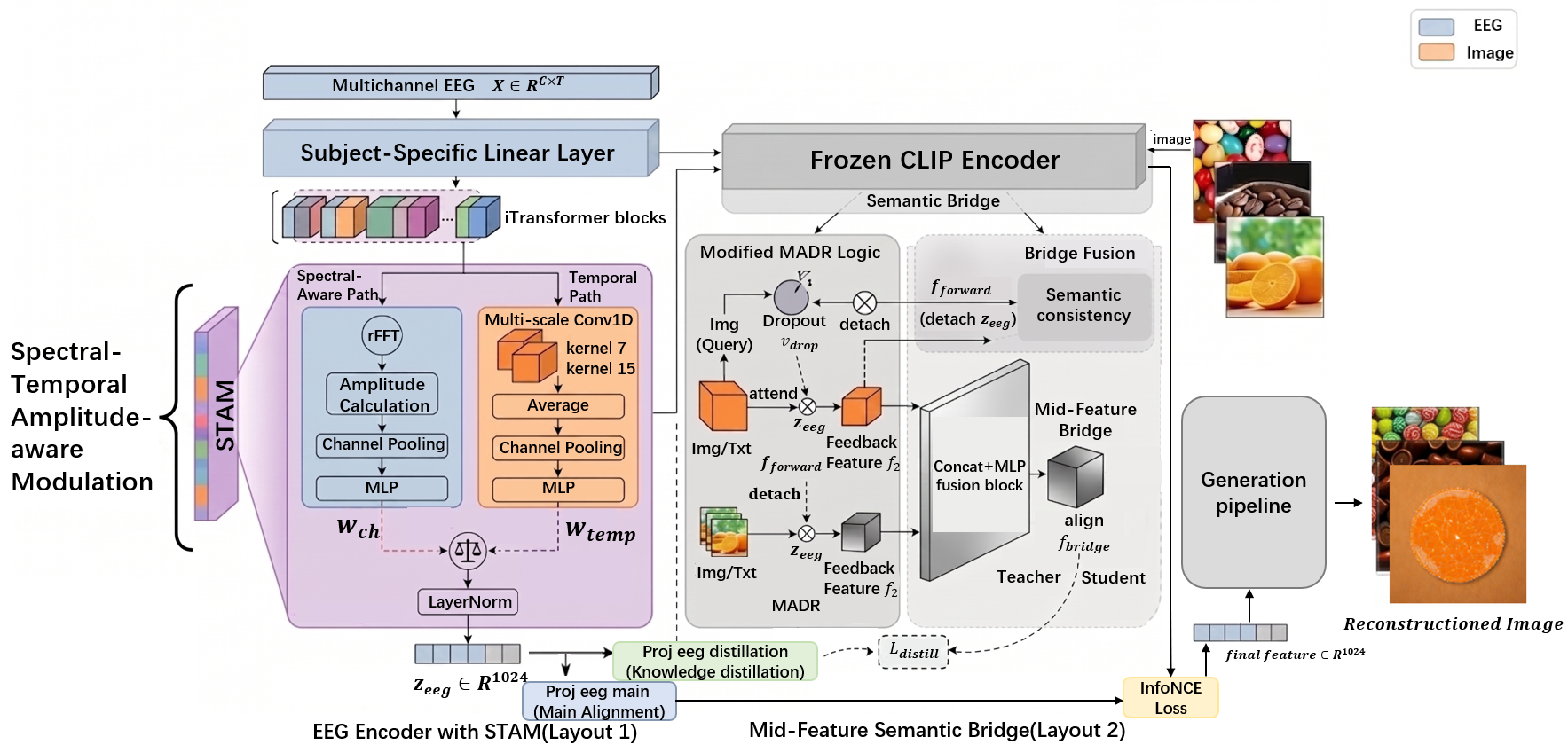}%
  }
  \caption{Overall framework of our proposed STAMBRIDGE model, illustrating the Spectral-Temporal Amplitude-aware Modulation (STAM) and the Mid-Feature Semantic Bridge (MFSB).}
  \label{fig:overall}
\end{figure*}

\section{Methods}
\label{sec:method}

The objective of STAMBRIDGE is to learn EEG representations that bridge low-SNR neural signals and vision--language embedding spaces (e.g., CLIP). Let $\mathcal{D}=\{(\mathbf{X}_i, \mathbf{v}_i, \mathbf{t}_i)\}_{i=1}^N$, where $\mathbf{X}_i \in \mathbb{R}^{C \times T}$ denotes EEG data and $\mathbf{v}_i, \mathbf{t}_i$ are pretrained semantic embeddings. To avoid suboptimal direct alignment, we adopt a decoupled architecture: STAM for spectral-temporal modulation and MFSB for semantic alignment.

\subsection{EEG Encoder with Spectral-Temporal Amplitude-aware Modulation}
\label{sec:method_encoder}
To effectively capture the spatio-temporal dynamics of neural signals and explicitly protect feature integrity, our encoding pipeline strictly follows a specific logic: Multichannel EEG $\rightarrow$ Subject-Specific Linear Layer $\rightarrow$ iTransformer blocks $\rightarrow$ STAM module. 

Raw EEG trials $\mathbf{X}_i \in \mathbb{R}^{C \times T}$ ($C=63$, $T=250$) are first processed by the Subject-Specific Linear Layer, which maps each subject's channel topology into a shared latent space via pre-calibrated projections to mitigate spatial variability. The adapted features are then processed by a stack of iTransformer blocks \cite{liu2024itransformer} to capture global long-range temporal dependencies.

The resulting sequence is then fed into the STAM. Recent studies indicate that under the extremely low-SNR conditions of rapid visual stimuli (RSVP), conventional spectral filtering—which often relies on hard frequency truncation followed by inverse Fourier transformation—may introduce temporal distortions that compromise transient neural dynamics \cite{xiong2025e2ivae, ma2026neurodecoder}.

To mitigate this issue, STAM adopts a spectral-informed channel modulation strategy rather than performing frequency-domain filtering. Instead, frequency-domain amplitudes are used as informative priors for adaptive channel modulation. Given the input tensor $\mathbf{X}_{in}\in\mathbb{R}^{B\times C\times T}$, STAM first computes the frequency-domain amplitude along the temporal axis using the real-valued Fast Fourier Transform (rFFT). Rather than applying a hard spectral mask, the mean amplitude serves as a global spectral descriptor that is fed into a two-layer bottleneck MLP (reduction ratio $r=8$) with Sigmoid activation, producing a sample-specific channel weighting matrix $\mathbf{W}_c\in(0,1)^{B\times C}$. The resulting weights are broadcast along the temporal dimension and multiplied element-wise with the original time-domain sequence, maintaining the output dimension of the spectral branch as $\mathbb{R}^{B\times C\times T}$.

Specifically, the temporal descriptor is computed via channel-wise mean pooling and passed through a two-layer MLP to generate a sample-specific temporal weight matrix $\mathbf{W}_t \in (0,1)^{B \times T}$, which adaptively scales the sequence along the time axis. 

Finally, let $\mathbf{X}_{spec} \in \mathbb{R}^{B \times C \times T}$ and $\mathbf{X}_{temp} \in \mathbb{R}^{B \times C \times T}$ denote the modulated outputs from the spectral and temporal branches, respectively. To dynamically balance frequency-aware features and time-domain transients, they are fused via a learnable weighted addition:
\begin{equation}
\mathbf{X}_{fused} = \lambda_1 \mathbf{X}_{spec} + \lambda_2 \mathbf{X}_{temp}, \quad \text{where} \;\; \boldsymbol{\lambda} = \text{Softmax}(\boldsymbol{\alpha}).
\end{equation}
Here, $\boldsymbol{\alpha} = [\alpha_1, \alpha_2]^\top \in \mathbb{R}^2$ is a learnable parameter that ensures the sum of fusion weights equals one ($\lambda_1 + \lambda_2 = 1$). To capture local spatial-temporal patterns before semantic projection, the fused output $\mathbf{X}_{fused}$ is further processed by a shallow convolutional patch embedding module (comprising depthwise/pointwise 2D convolutions and average pooling). It is then flattened and passed through a projection head—comprising sequential Linear, GELU, Dropout, and Layer Normalization layers—to match the $d=1024$ dimensional semantic space, yielding $\mathbf{z}_{eeg} \in \mathbb{R}^{d}$ as the primary neural representation. Because STAM explicitly avoids hard masking, it preserves critical time-domain transients, a necessity for future tasks requiring regional and spatial semantic control \cite{xiang2026regional}.

\subsection{Mid-Feature Semantic Bridge}
\label{sec:method_bridge}
To mitigate the instability inherent in direct one-step mapping between $\mathbf{z}_{eeg}$ and visual semantic spaces, we construct a softened intermediate target using a novel Multi-modal Adaptive Directional Routing (MADR) attention mechanism. Inspired by the success of cross-modal attention in establishing semantic consistency for general image--text matching \cite{xu2020crossmodal}, this design utilizes directed attention to explicitly model how stable semantic modalities interpret the current dynamic neural state, providing a context-aware target for refinement.

\subsubsection{Multi-modal Adaptive Directional Routing (MADR)}
In the STAMBRIDGE framework, we intentionally utilize only the high-level semantic pipeline from the ATM framework and omit the low-level pipeline to prevent overfitting. Consequently, the input visual and textual representations are highly condensed single global CLIP embeddings (i.e., sequence length = 1) rather than token sequences. 

As applying standard self-attention within a single token sequence is trivial, our MADR mechanism is mathematically formulated to perform attention across the modality dimension. Specifically, given a batch size $B$, number of attention heads $H$, and head dimension $d$, the EEG query ($\mathbf{Q}$) has an exact tensor shape of $\mathbb{R}^{B \times H \times 1 \times d}$. The key ($\mathbf{K}$) and value ($\mathbf{V}$), which are formed by concatenating the visual and textual embeddings, possess a shape of $\mathbb{R}^{B \times H \times 2 \times d}$. The resulting attention scores form a matrix of $\mathbb{R}^{B \times H \times 1 \times 2}$. The Softmax operation is applied over the last dimension (size=2), acting as a dynamic cross-modal router.

Standard cross-attention typically treats modalities uniformly. In contrast, MADR dynamically estimates a routing weight matrix to balance visual and textual semantics according to the current state of the neural embedding.

Given a query $\mathbf{Q}$ from one modality and a set of keys/values $\{\mathbf{K}_m, \mathbf{V}_m\}_{m=1}^M$ from $M$ target modalities (e.g., image and text), MADR first computes a head-aware routing probability matrix $\mathbf{W}_{route} \in \mathbb{R}^{H \times M}$ using a routing multi-layer perceptron (MLP), where $H$ denotes the number of attention heads:
\begin{equation}
\mathbf{W}_{route} = \text{Softmax}(\text{MLP}_{route}(\text{Pool}(\mathbf{Q}))).
\end{equation}
Here, $\text{Pool}(\cdot)$ denotes mean pooling over the token dimension. Specifically, the $\text{MLP}_{route}$ consists of two linear layers separated by a ReLU activation (with a hidden dimension of 128). For each of the $H$ attention heads, the MLP generates an $M$-dimensional probability distribution ($M=2$ for image and text modalities), yielding the $H \times M$ routing space. This fine-grained design allows each individual head to independently balance its focus between visual and textual semantics. Note that the target visual and textual prototypes ($\{\mathbf{K}_m, \mathbf{V}_m\}$) are pre-extracted using a strictly frozen OpenCLIP ViT-H/14 encoder (pre-trained on the LAION-2B dataset) to ensure a stable reference manifold. To prevent routing collapse during early training, the final linear projection bias is initialized to zero, ensuring an approximately uniform initial routing distribution (i.e., $W_{\mathrm{route}, i, m} \approx \frac{1}{M}$). The resulting routing matrix independently modulates the cross-attention outputs of each individual head (rather than globally per module) before aggregation over modalities. For the $i$-th head, the routed attention output is defined as:
\begin{equation}
\mathbf{H}_i = \sum_{m=1}^{M} W_{route, i, m} \cdot \text{Softmax}\left(\frac{\mathbf{Q}_i \mathbf{K}_{m,i}^\top}{\sqrt{d_k}}\right)\mathbf{V}_{m,i},
\end{equation}
where $W_{route, i, m}$ represents the scalar routing weight corresponding to the $i$-th head and $m$-th modality (i.e., the $(i, m)$-th element of $\mathbf{W}_{route}$). Additionally, $\mathbf{Q}_i$, $\mathbf{K}_{m,i}$, and $\mathbf{V}_{m,i}$ denote the projected query, key, and value matrices specific to the $i$-th head, and $d_k$ is the dimensionality of the head-specific key/query vectors.

\subsubsection{Semantic Interaction and Staged Distillation}
We employ MADR to construct the regularized Mid-Feature space. Specifically, the neural representation $\mathbf{z}_{eeg}$ acts as the query to attend to frozen visual and textual prototypes, which serve as the keys and values. This attention mechanism explicitly captures how the current dynamic neural state aligns with and retrieves information from stable semantic modalities. Let $\mathbf{h}_{attn} \in \mathbb{R}^d$ denote the final aggregated cross-attention output. While MADR aggregates both visual and textual concepts to form a rich semantic context, the final intermediate target must be specifically anchored to the visual modality to optimize for downstream visual decoding tasks. To robustify this representation against low-SNR artifacts and prevent deterministic collapse during training, $\mathbf{h}_{attn}$ is concatenated with stochastically dropped-out visual features and passed through a fusion network. Formally, the regularized Mid-Feature target, denoted as $\mathbf{f}_{bridge}$, is formulated as:
\begin{equation}
\mathbf{f}_{bridge} = \mathcal{N}\Big( \text{MLP}_{fuse}\big( [ \mathbf{h}_{attn} \parallel \text{Dropout}(\mathbf{v}, p) ] \big) \Big),
\end{equation}
where $[\cdot \parallel \cdot]$ denotes concatenation along the feature dimension, $\mathbf{v}$ is the target visual prototype, and $\text{Dropout}(\cdot, p)$ applies stochastic masking with probability $p$ (empirically set to $0.4$). The fusion network $\text{MLP}_{fuse}$ consists of Layer Normalization, linear projections, and GELU activations, while $\mathcal{N}(\cdot)$ represents strictly enforced L2-normalization. The resulting output $\mathbf{f}_{bridge}$ serves as the context-aware, softened target for the subsequent staged distillation.

Unlike raw visual embeddings, $\mathbf{f}_{bridge}$ is deeply conditioned on the neural context and therefore provides a smoother optimization target. During training, we adopt a \textbf{staged distillation} strategy formulated as a multi-task objective. 
First, to maintain global semantic consistency, the primary EEG embedding $\mathbf{z}_{eeg}$ is directly aligned with the visual semantic manifold via the InfoNCE loss \cite{oord2018representation}:
\begin{equation}
\mathcal{L}_{main} = \mathcal{L}_{\text{InfoNCE}}(\mathbf{z}_{eeg}, \mathbf{v}).
\end{equation}
Simultaneously, the intermediate bridge is anchored to the target visual semantic manifold:
\begin{equation}
\mathcal{L}_{bridge} = \mathcal{L}_{\text{InfoNCE}}(\mathbf{f}_{bridge}, \mathbf{v}).
\end{equation}
Finally, an auxiliary projection of the EEG embedding is distilled toward this aligned intermediate representation. To stabilize optimization and avoid degenerate coupling, $\mathbf{f}_{bridge}$ is detached from the computation graph during distillation:
\begin{equation}
\mathcal{L}_{distill} = \mathcal{L}_{\text{InfoNCE}}(\text{Proj}(\mathbf{z}_{eeg}), \mathbf{f}_{bridge}^{\text{detach}}).
\end{equation}
Here, $\text{Proj}(\cdot)$ denotes an auxiliary projection head with stabilized residual initialization and orthogonal weights to prevent gradient interference. The total objective integrates these components:
\begin{equation}
\mathcal{L}_{total} = \lambda_0 \mathcal{L}_{main} + \lambda_1 \mathcal{L}_{bridge} + \lambda_2 \mathcal{L}_{distill}.
\end{equation}
The latent representations ($\mathbf{f}_{bridge}, \mathbf{v}, \mathbf{z}_{eeg}$) are strictly $\ell_2$-normalized within the unified $d=1024$ space. The InfoNCE objectives follow a symmetric formulation parameterized by a learnable temperature $\tau$. To ensure a smooth transition from global alignment to fine-grained distillation, we empirically set $\lambda_0 = 0.99, \lambda_1 = 0.5$, while linearly warming up $\lambda_2$ from $0.0$ to $0.2$ across epochs. Rather than a strictly sequential two-stage process or a simultaneous joint optimization (which suffers from early gradient conflicts), this delayed-activation multi-objective optimization explicitly acts as a buffer against the modality gap. Furthermore, to clearly distinguish STAMBRIDGE from existing frameworks like Neural-MCRL, our model fundamentally diverges in its architecture: it employs a Subject-Specific Linear Layer for spatial variance and exclusively utilizes a high-level semantic pipeline to prevent overfitting, ensuring that the main feature extraction path is robustly optimized.

\section{Experiments}
\label{sec:experiments}

\subsection{Experimental Setup}
\paragraph{Dataset and Preprocessing}
We evaluate our framework on the THINGS-EEG dataset \cite{gifford2022large}, a large-scale benchmark containing EEG recordings from 10 subjects during the RSVP of natural images sourced from the THINGS database \cite{hebart2019things}. The dataset comprises 16,540 training image concepts and 200 distinct test concepts. Following standard protocols, the continuous EEG data is segmented into trials from 0 to 1000 ms relative to stimulus onset, baseline-corrected, and downsampled to 250 Hz, resulting in 63-channel sequence inputs.

\paragraph{Implementation Details}
Experiments were conducted on a single NVIDIA RTX 3090 GPU. For evaluation, we strictly follow a high-level-only two-stage generation pipeline \cite{li2024visual} utilizing a 50-step Diffusion Prior and a 4-step IP-Adapter \cite{ye2023ip} conditioned Stable Diffusion model. To prevent any risk of target leakage, no ground-truth image/text features or MFSB bridge features are utilized at test time; the generative prior is conditioned exclusively on projected raw EEG representations. Complete training hyperparameters and the step-by-step inference pseudocode are detailed in Algorithm S1 of the Supplementary Material.

\paragraph{Evaluation Metrics}
To evaluate cross-modal semantic alignment, we report Top-1 and Top-5 accuracy in a 200-way zero-shot retrieval setting. For the downstream qualitative image generation task, we quantitatively assess semantic consistency using standard metrics. To highlight semantic alignment, we report high-level metrics (AlexNet(2), AlexNet(5), InceptionV3, and CLIP similarity) in the main text. Additional structural metrics (PixCorr, SSIM) and deep clustering features (EffNet-B, SwAV) are comprehensively detailed in the Supplementary Material.

\subsection{Zero-Shot Image Retrieval Results}
We compare STAMBRIDGE against recent EEG-based visual decoding models, including BraVL \cite{du2023decoding}, NICE \cite{song2024decoding}, MB2C \cite{wei2024mb2c}, ATM \cite{li2024visual}, and Neural-MCRL \cite{li2025neural}. Table \ref{tab:main_retrieval} presents the \textbf{average} 200-way zero-shot retrieval accuracy across all 10 subjects. \textbf{The comprehensive per-subject retrieval results are fully detailed in Supplementary Table S3.}

\begin{table}[!t]
\centering
\caption{Overall 200-Way Zero-Shot Retrieval Accuracy (\%) on the THINGS-EEG Dataset.}
\label{tab:main_retrieval}
\resizebox{\columnwidth}{!}{
\begin{tabular}{lcc}
\toprule
\textbf{Method} & \textbf{Top-1 (\%)} & \textbf{Top-5 (\%)} \\
\midrule
BraVL [21] & 5.81 $\pm$ 1.42  & 16.84 $\pm$ 3.32 \\
NICE [20] & 14.80 $\pm$ 3.21 & 40.10 $\pm$ 6.91 \\
MB2C [36] & 27.70 $\pm$ 5.89 & 58.95 $\pm$ 7.24 \\
ATM [12] & 29.35 $\pm$ 6.37 & 56.35 $\pm$ 8.97 \\
Neural-MCRL [6] & 31.05 $\pm$ 5.71 & 59.70 $\pm$ 7.36 \\
\midrule
\textbf{STAMBRIDGE (Ours)} & \textbf{34.50 $\pm$ 6.14$^{***}$} & \textbf{65.95 $\pm$ 5.85$^{***}$} \\
\bottomrule
\end{tabular}
}
\vspace{2pt}
\raggedright \scriptsize 
\textit{Note:} Results are reported as mean $\pm$ std across 10 subjects. Detailed per-subject results and statistical testing are provided in Supplementary Tables.
\end{table}

As shown in Table \ref{tab:main_retrieval}, STAMBRIDGE establishes a strong result on this benchmark. Compared with Neural-MCRL, our model achieves statistically significant improvements in both Top-1 and Top-5 retrieval accuracy. Although the absolute numerical improvement in Top-1 accuracy is moderate (+3.45\%), statistical analysis demonstrates that the gain is highly consistent across all ten subjects (paired t-test, $p<0.001$, Cohen's $d=1.64$). This indicates that the observed improvements are robust across individuals rather than resulting from random variation. Detailed statistical analyses and \textbf{the complete per-subject breakdowns} are provided in Supplementary \textbf{Tables S2 and S3}.

\subsection{Model-Agnostic Generality of Semantic Bridge}
To explicitly validate the plug-and-play nature of the proposed Mid-Feature Semantic Bridge, we integrate it as an independent module into three distinct baseline architectures: NICE \cite{song2024decoding}, ATM \cite{li2024visual}, and Neural-MCRL \cite{li2025neural}. We evaluate the average 200-way zero-shot retrieval accuracy across all subjects under identical experimental settings.

\begin{table}[!t]
\centering
\caption{Cross-Model Generalization of the Mid-Feature Semantic Bridge}
\label{tab:plugin_comparison}
\resizebox{\columnwidth}{!}{%
\begin{tabular}{lcc}
\toprule
\textbf{Model Configuration} & \textbf{Top-1 (\%)} & \textbf{Top-5 (\%)} \\
\midrule
NICE (Baseline)          & 14.80 $\pm$ 3.21 & 40.10 $\pm$ 6.91 \\
\textbf{NICE + MFSB} & \textbf{15.40 $\pm$ 3.89} & \textbf{42.70 $\pm$ 4.82}$^{***}$ \\
\midrule
ATM (Baseline)           & 29.35 $\pm$ 6.37 & 56.35 $\pm$ 8.97 \\
\textbf{ATM + MFSB}   & \textbf{30.55 $\pm$ 5.11} & \textbf{61.35 $\pm$ 7.58}$^{*}$ \\
\midrule
Neural-MCRL (Baseline)   & 31.05 $\pm$ 5.71 & 59.70 $\pm$ 7.36 \\
\textbf{Neural-MCRL + MFSB} & \textbf{32.75 $\pm$ 4.91}$^{*}$ & \textbf{62.70 $\pm$ 5.87}$^{*}$ \\
\bottomrule
\end{tabular}%
}
\vspace{2pt}
\raggedright \scriptsize 
\textit{Note:} Results are mean $\pm$ std across 10 subjects. Asterisks denote statistical significance compared to the baseline ($^{*}p < 0.05, ^{***}p < 0.001$). Detailed testing is provided in Supplementary Table S2.
\end{table}

As shown in Table \ref{tab:plugin_comparison}, integrating the proposed semantic bridge consistently improves retrieval performance across all evaluated backbone architectures. In particular, statistically significant improvements are observed for Top-5 retrieval ($p<0.05$ or $p<0.001$), demonstrating that the proposed bridge generalizes well beyond our own architecture. By providing a regularized intermediate target, the bridge facilitates a more stable cross-modal alignment trajectory regardless of the underlying EEG feature extractor.

\subsection{Ablation Studies: Disentangling Module Contribution}
To systematically evaluate the individual contributions of the STAM and MFSB within our full framework, we conducted ablation studies under identical training conditions. The average retrieval results are summarized in Table \ref{tab:ablation}.

\begin{table}[!t]
\centering
\caption{Extensive Ablation Studies on Core Components, Training Strategies, Shortcut Control, Fusion Mechanisms, and Modulation Strategies.}
\label{tab:ablation}
\resizebox{\columnwidth}{!}{
\begin{tabular}{lcc}
\toprule
\textbf{Configuration} & \textbf{Top-1 Acc. (\%)} & \textbf{Top-5 Acc. (\%)} \\
\midrule
\multicolumn{3}{l}{\textit{(a) Core Component Ablation}} \\
Baseline (Neural-MCRL) & 31.05 $\pm$ 5.71 & 59.70 $\pm$ 7.36 \\
+ MFSB Only            & 32.75 $\pm$ 4.91 & 62.70 $\pm$ 5.87$^{*}$ \\
+ STAM Only            & 33.95 $\pm$ 5.19$^{***}$ & 63.60 $\pm$ 6.66$^{**}$ \\
\textbf{STAMBRIDGE (Full)} & \textbf{34.50 $\pm$ 6.14$^{***}$} & \textbf{65.95 $\pm$ 5.85$^{***}$} \\
\midrule
\multicolumn{3}{l}{\textit{(b) Training Strategy Optimization}} \\
Direct InfoNCE         & 33.95 $\pm$ 5.19 & 63.60 $\pm$ 6.66 \\
Truly Sequential       & 33.25 $\pm$ 4.97 & 64.70 $\pm$ 6.84 \\
Simultaneous Joint     & 33.60 $\pm$ 6.36 & \textbf{66.20 $\pm$ 7.68} \\
\textbf{Warm-up (Ours)}& \textbf{34.50 $\pm$ 6.14} & 65.95 $\pm$ 5.85 \\
\midrule
\multicolumn{3}{l}{\textit{(c) MFSB Visual Drop Rate}} \\
Drop = 1.0 (Pure EEG)  & 32.65 $\pm$ 7.02 & 62.05 $\pm$ 8.16 \\
Drop = 0.8             & 32.95 $\pm$ 6.92 & 64.45 $\pm$ 7.88 \\
Drop = 0.6             & 33.30 $\pm$ 7.27 & 64.25 $\pm$ 8.69 \\
\textbf{Drop = 0.4 (Ours)} & \textbf{34.50 $\pm$ 6.14} & \textbf{65.95 $\pm$ 5.85} \\
\midrule
\multicolumn{3}{l}{\textit{(d) Cross-Modal Fusion Mechanisms}} \\
Standard Cross-Attention & 33.65 $\pm$ 5.52 & 65.75 $\pm$ 7.37 \\
Simple Weighted Avg      & 33.70 $\pm$ 7.33 & 65.10 $\pm$ 7.78 \\
Concatenation Fusion     & 34.15 $\pm$ 7.08 & \textbf{66.20 $\pm$ 7.35} \\
\textbf{MADR (Ours)}     & \textbf{34.50 $\pm$ 6.14} & 65.95 $\pm$ 5.85 \\
\midrule
\multicolumn{3}{l}{\textit{(e) Spectral-Temporal Modulation Mechanisms}} \\
Hard Spectral Masking    & 33.45 $\pm$ 6.48 & 65.65 $\pm$ 6.61 \\
Tapered Filtering        & 33.90 $\pm$ 5.48 & \textbf{66.65 $\pm$ 6.25} \\
Standard Channel Attn.   & 33.50 $\pm$ 5.80 & 65.45 $\pm$ 6.86 \\
Time-Domain Gating       & 34.10 $\pm$ 4.81 & 66.45 $\pm$ 4.95 \\
\textbf{STAM Module (Ours)} & \textbf{34.50 $\pm$ 6.14} & 65.95 $\pm$ 5.85 \\
\bottomrule
\end{tabular}
}
\vspace{2pt}
\raggedright \scriptsize 
\textit{Note:} Accuracy is reported as mean $\pm$ std across 10 subjects. Asterisks denote statistical significance compared to the Baseline ($^{*}p < 0.05, ^{**}p < 0.01, ^{***}p < 0.001$). Comprehensive per-subject breakdowns are detailed in Supplementary Table S4.
\end{table}

As shown in Table \ref{tab:ablation}, we systematically evaluate the individual contributions and design choices of our framework. 

Table \ref{tab:ablation}(a) reveals the complementary roles of our core components: STAM primarily drives discriminative representation learning, while MFSB further stabilizes the semantic alignment (improving Top-5 retrieval). To rigorously address optimization conflicts, Table \ref{tab:ablation}(b) compares training strategies. While simultaneous joint optimization yields a marginally higher Top-5 score, our warm-up optimization provides the best Top-1 accuracy (34.50\%) and balances the optimization trajectory, verifying that delayed activation prevents early gradient conflicts while allowing dynamic refinement. 

Furthermore, to rule out the possibility that the MFSB exploits target image dimensions as an informational shortcut, Table \ref{tab:ablation}(c) demonstrates that even when the visual target is completely dropped during training (Drop = 1.0), the pure EEG-driven bridge sustains a competitive Top-1 accuracy of 32.65\%. This robustly confirms that the retained visual dimensions in our standard setting act merely as routing guides rather than shortcuts. Finally, Table \ref{tab:ablation}(d) validates the MADR module. Although simple concatenation achieves a slightly higher Top-5 accuracy, MADR consistently delivers the best Top-1 retrieval performance, demonstrating that its cross-modal directional routing is highly effective for fine-grained discrimination.

Finally, Table \ref{tab:ablation}(e) provides a quantitative comparison of alternative temporal and spectral modulation baselines, which directly corroborates our signal-level visual analysis. While tapered filtering and time-domain gating achieve marginally higher Top-5 accuracies, our STAM module achieves the highest Top-1 accuracy (34.50\%). This confirms that explicitly preserving frequency-aware transients without time-domain masking artifacts translates directly into superior fine-grained semantic discrimination.

\subsection{Qualitative Validation and Visualization}
\label{sec:qualitative_results}
To further validate the structural integrity of the learned EEG embeddings, we perform visual reconstruction and latent space analysis.

\paragraph{Signal-Level Validation of STAM}
To empirically validate the time-domain integrity preserved by STAM, we compared its waveform-level outputs against traditional frequency-domain filters (which often introduce severe ringing artifacts) and temporal/channel gating mechanisms (which tend to over-smooth the signal). As detailed in the Supplementary Material (see \textbf{Figs. S24--S33} in the Supplementary Material), STAM explicitly preserves stimulus-locked high-frequency neural transients without inducing artificial oscillations, confirming its structural superiority for latency-sensitive RSVP paradigms.

\paragraph{Neurophysiological Interpretability via Spatial Attention}
To verify reliance on biologically meaningful signals, we analyzed the EEG encoder's spatial attention weights (see \textbf{Fig. S3} in the Supplementary Material). The activation maps across all 10 subjects and their grand average consistently highlight the occipital and parietal regions, aligning with the neurophysiological basis of human visual processing. This consistency validates that our Subject-Specific Linear Layer and STAM effectively mitigate spatial variance while preserving task-relevant cues before cross-modal alignment.

\paragraph{Latent Alignment and Retrieval Visualization}
To intuitively understand the effectiveness of our MFSB, we visualize the learned feature representations using t-SNE (see \textbf{Fig. S2} in the Supplementary Material). The visualization reveals that the EEG features exhibit clustering patterns corresponding to the underlying visual semantics, indicating improved intra-class compactness and inter-class separability. Furthermore, as demonstrated by the qualitative zero-shot retrieval results (\textbf{Fig. S1}), given an unseen EEG query, the model retrieves images that generally share the same high-level semantic category. Even when the Top-1 result is not the exact target image, the retrieved candidates remain semantically consistent, often preserving the shape, color palette, and object category. These observations validate that the proposed MFSB effectively constrains neural embeddings within a semantically structured latent manifold.

\paragraph{Semantic Image Reconstruction via Generative Prior}
As established, we treat the generative pipeline strictly as a qualitative validation tool for our latent space, utilizing exclusively the high-level semantic pipeline to prevent spatial overfitting. Table \ref{tab:main_generation} presents a concise quantitative comparison of generative performance, focusing strictly on high-level semantic indicators. Our model yields constrained performance on low-level structural metrics (e.g., SSIM) compared to ATM \cite{li2024visual}. This is expected, as we explicitly omit the spatial regression branch. However, the true measure of latent cognitive alignment relies on high-level semantics. As shown in Table \ref{tab:main_generation}, STAMBRIDGE demonstrates a substantial lead across major deep semantic metrics when compared to the re-implemented ATM across all 10 subjects and the single-modality baseline CognitionCapturer \cite{zhang2025cognitioncapturer}. The comprehensive per-subject quantitative breakdown and qualitative visual reconstructions are fully documented in the Supplementary Material (see \textbf{Table S1} and \textbf{Figs. S4--S23}).

\paragraph{Qualitative Analysis of Successes and Failures}
As visualized in the comprehensive generation results (see \textbf{Figs. S4--S13} in the Supplementary Material), the qualitative performance further reveals the distinct characteristics of our learned representations. In highly successful cases, STAMBRIDGE perfectly captures overarching semantic categories—such as food (e.g., coffee beans, oranges), animals, and basic objects (e.g., spoons, basketballs)—preserving both color palettes and global contours. This semantic consistency confirms that the STAM module effectively preserves stimulus-locked transients, while the MFSB successfully anchors these noisy signals to the proper CLIP semantic manifold.

However, qualitative analysis also reveals specific failure modes intrinsically tied to the nature of EEG signals. The most common failure is \textbf{semantic deviation within sub-categories}. For example, the model occasionally reconstructs a cat as a dog, a rhinoceros as a deer, or a gorilla as a bear. Because these distinct species share proximate distances in the CLIP latent space, the extremely low SNR of the RSVP paradigm makes it challenging to decode fine-grained semantic boundaries from brief neural transients. Furthermore, instances of \textbf{missing details} or \textbf{structural misalignment} (e.g., reconstructing a complex vehicle as a general landscape or room) occasionally occur. This is an expected trade-off: as discussed, we explicitly omitted the deterministic low-level pixel regression pathway to prevent spatial mean collapse caused by the severe Volume Conduction Effect of EEG \cite{li2025tale}. Consequently, while the framework achieves robust high-level cognitive alignment, it struggles to reconstruct deterministic 2D spatial layouts for complex artificial structures.

\begin{table}[!t]
\centering
\caption{Quantitative Comparison of Generative Performance Focusing on Semantic Alignment.}
\label{tab:main_generation}
\resizebox{\columnwidth}{!}{%
\begin{tabular}{lcccc}
\toprule
\textbf{Method} & \textbf{AlexNet(2) $\uparrow$} & \textbf{AlexNet(5) $\uparrow$} & \textbf{Inception $\uparrow$} & \textbf{CLIP $\uparrow$} \\
\midrule
CognitionCapturer [37] & \textbf{0.813 $\pm$ 0.025} & 0.671 $\pm$ 0.038 & 0.664 $\pm$ 0.043 & 0.705 $\pm$ 0.041 \\
ATM [12] (Re-impl.) & 0.783 $\pm$ 0.011 & \textbf{0.860 $\pm$ 0.013} & 0.704 $\pm$ 0.029 & 0.719 $\pm$ 0.028 \\
\textbf{STAMBRIDGE (Ours)} & {0.784 $\pm$ 0.022} & \textbf{0.860 $\pm$ 0.023} & \textbf{0.717 $\pm$ 0.036} & \textbf{0.730 $\pm$ 0.038} \\
\bottomrule
\end{tabular}%
}
\vspace{2pt}
\raggedright \scriptsize 
\textit{Note:} Results are reported as mean $\pm$ std across 10 subjects. Complete evaluation metrics (including spatial/structural metrics) are detailed in Supplementary Table S1.
\end{table}

\section{Discussion}
\label{sec:discussion}

Unlike fMRI, EEG suffers from severe spatial smearing, making direct pixel-level reconstruction highly challenging. To avoid Spatial Mean Collapse, our pipeline prioritizes preserving the neural backbone using a Subject-Specific Linear Layer, an iTransformer-based encoder, and STAM, ensuring that frequency-aware transients remain available for semantic decoding. Crucially, by intentionally utilizing only the high-level semantic branch of ATM and omitting its low-level reconstruction branch, we mitigate potential overfitting to noisy or unreliable low-level spatial details. Furthermore, we achieve robust cross-modal alignment through internal staged distillation via the MFSB, avoiding the additional computational and deployment overhead associated with external dependencies such as RAG or MLLMs.

While STAMBRIDGE yields significant stability advantages, a notable limitation stems from CLIP-based semantic alignment. Since the RSVP paradigm mainly evokes coarse categorical representations in EEG, mapping these signals into dense CLIP semantic clusters makes the model susceptible to fine-grained misclassifications (e.g., confusing related concepts like "cat" and "dog"). Future work will explore adapting the model-agnostic MFSB module to more biologically plausible visual encoders (requiring only minor dimensional adjustments to the initial projection layers). Additionally, we will investigate accelerated diffusion decoding and domain adaptation strategies to enable more efficient and subject-independent EEG visual decoding.

\section{Conclusion}
\label{sec:conclusion}
STAMBRIDGE is a framework for EEG-based visual decoding that moves beyond brittle one-step alignment. Its Spectral-Temporal Amplitude-aware Modulation (STAM) uses soft channel weighting and multi-scale temporal convolutions to preserve frequency-aware transients without time-domain masking artifacts. Building on this, the Mid-Feature Semantic Bridge (MFSB) constructs a regularized intermediate space and eases cross-modal alignment through staged distillation.

Empirically, STAMBRIDGE achieves 34.50\% Top-1 and 65.95\% Top-5 accuracy on the THINGS-EEG benchmark. Ablation and plug-in results further show that MFSB is a model-agnostic module that improves optimization stability and candidate coverage. Although cross-subject generalization remains challenging, the proposed framework provides a solid foundation for semantically aligned EEG representation learning.

\bibliographystyle{unsrt}
\bibliography{references}

@article{xu2020crossmodal,
  title={Cross-modal attention with semantic consistence for image--text matching},
  author={Xu, Xing and Wang, Tan and Yang, Yang and Zuo, Lin and Shen, Fumin and Shen, Heng Tao},
  journal={IEEE transactions on neural networks and learning systems},
  volume={31},
  number={12},
  pages={5412--5425},
  year={2020},
  publisher={IEEE}
}

@article{li2024visual,
  title={Visual Decoding and Reconstruction via EEG Embeddings with Guided Diffusion},
  author={Li, Dongyang and Wei, Chen and Li, Shiying and Zou, Jiachen and Liu, Quanying},
  journal={Advances in Neural Information Processing Systems},
  volume={37},
  pages={102822--102864},
  year={2024}
}

@inproceedings{li2025neural,
  title={Neural-MCRL: Neural multimodal contrastive representation learning for EEG-based visual decoding},
  author={Li, Yueyang and Kang, Zijian and Gong, Shengyu and Dong, Wenhao and Zeng, Weiming and Yan, Hongjie and Siok, Wai Ting and Wang, Nizhuan},
  booktitle={2025 IEEE International Conference on Multimedia and Expo (ICME)},
  pages={1--6},
  year={2025},
  organization={IEEE}
}

@inproceedings{zhang2025cognitioncapturer,
  title={Cognitioncapturer: Decoding visual stimuli from human eeg signal with multimodal information},
  author={Zhang, Kaifan and He, Lihuo and Jiang, Xin and Lu, Wen and Wang, Di and Gao, Xinbo},
  booktitle={Proceedings of the AAAI Conference on Artificial Intelligence},
  volume={39},
  number={13},
  pages={14486--14493},
  year={2025}
}

@article{gifford2022large,
  title={A large and rich EEG dataset for modeling human visual object recognition},
  author={Gifford, Alessandro T and Dwivedi, Kshitij and Roig, Gemma and Cichy, Radoslaw M},
  journal={NeuroImage},
  volume={264},
  pages={119754},
  year={2022},
  publisher={Elsevier}
}

@article{hebart2019things,
  title={THINGS: A database of 1,854 object concepts and more than 26,000 naturalistic object images},
  author={Hebart, Martin N and Dickter, Adam H and Kidder, Alexis and Kwok, Wan Y and Corriveau, Anna and Van Wicklin, Caitlin and Baker, Chris I},
  journal={PloS one},
  volume={14},
  number={10},
  pages={e0223792},
  year={2019},
  publisher={Public Library of Science San Francisco, CA USA}
}

@inproceedings{radford2021learning,
  title={Learning transferable visual models from natural language supervision},
  author={Radford, Alec and Kim, Jong Wook and Hallacy, Chris and Ramesh, Aditya and Goh, Gabriel and Agarwal, Sandhini and Sastry, Girish and Askell, Amanda and Mishkin, Pamela and Clark, Jack and others},
  booktitle={International conference on machine learning},
  pages={8748--8763},
  year={2021},
  organization={PmLR}
}

@inproceedings{liu2024itransformer,
  title={itransformer: Inverted transformers are effective for time series forecasting},
  author={Liu, Yong and Hu, Tengge and Zhang, Haoran and Wu, Haixu and Wang, Shiyu and Ma, Lintao and Long, Mingsheng},
  booktitle={International conference on learning representations},
  volume={2024},
  pages={11116--11140},
  year={2024}
}

@article{oord2018representation,
  title={Representation learning with contrastive predictive coding},
  author={Oord, Aaron van den and Li, Yazhe and Vinyals, Oriol},
  journal={arXiv preprint arXiv:1807.03748},
  year={2018}
}

@inproceedings{song2024decoding,
  title={Decoding natural images from eeg for object recognition},
  author={Song, Yonghao and Liu, Bingchuan and Li, Xiang and Shi, Nanlin and Wang, Yijun and Gao, Xiaorong},
  booktitle={International conference on learning representations},
  volume={2024},
  pages={47648--47665},
  year={2024}
}

@article{du2023decoding,
  title={Decoding visual neural representations by multimodal learning of brain-visual-linguistic features},
  author={Du, Changde and Fu, Kaicheng and Li, Jinpeng and He, Huiguang},
  journal={IEEE Transactions on Pattern Analysis and Machine Intelligence},
  volume={45},
  number={9},
  pages={10760--10777},
  year={2023},
  publisher={IEEE}
}

@inproceedings{rombach2022high,
  title={High-resolution image synthesis with latent diffusion models},
  author={Rombach, Robin and Blattmann, Andreas and Lorenz, Dominik and Esser, Patrick and Ommer, Bj{\"o}rn},
  booktitle={Proceedings of the IEEE/CVF conference on computer vision and pattern recognition},
  pages={10684--10695},
  year={2022}
}

@article{ye2023ip,
  title={Ip-adapter: Text compatible image prompt adapter for text-to-image diffusion models},
  author={Ye, Hu and Zhang, Jun and Liu, Sibo and Han, Xiao and Yang, Wei},
  journal={arXiv preprint arXiv:2308.06721},
  year={2023}
}

@article{song2022eeg,
  title={EEG conformer: Convolutional transformer for EEG decoding and visualization},
  author={Song, Yonghao and Zheng, Qingqing and Liu, Bingchuan and Gao, Xiaorong},
  journal={IEEE Transactions on Neural Systems and Rehabilitation Engineering},
  volume={31},
  pages={710--719},
  year={2022},
  publisher={IEEE}
}

@article{wu2022transfer,
  title={Transfer learning for motor imagery based brain--computer interfaces: A tutorial},
  author={Wu, Dongrui and Jiang, Xue and Peng, Ruimin},
  journal={Neural Networks},
  volume={153},
  pages={235--253},
  year={2022},
  publisher={Elsevier}
}

@article{li2019domain,
  title={Domain adaptation for EEG emotion recognition based on latent representation similarity},
  author={Li, Jinpeng and Qiu, Shuang and Du, Changde and Wang, Yixin and He, Huiguang},
  journal={IEEE Transactions on Cognitive and Developmental Systems},
  volume={12},
  number={2},
  pages={344--353},
  year={2019},
  publisher={IEEE}
}

@article{li2025tale,
  title={A Tale of Single-channel Electroencephalogram: Devices, Datasets, Signal Processing, Applications, and Future Directions},
  author={Li, Yueyang and Zeng, Weiming and Dong, Wenhao and Han, Di and Chen, Lei and Chen, Hongyu and Kang, Zijian and Gong, Shengyu and Yan, Hongjie and Siok, Wai Ting and others},
  journal={IEEE Transactions on Instrumentation and Measurement},
  pages={1--20},
  year={2025},
  publisher={IEEE}
}

@article{widmann2012filter,
  title={Filter effects and filter artifacts in the analysis of electrophysiological data},
  author={Widmann, Andreas and Schr{\"o}ger, Erich},
  journal={Frontiers in psychology},
  volume={3},
  pages={233},
  year={2012},
  publisher={Frontiers Research Foundation}
}

@article{widmann2015digital,
  title={Digital filter design for electrophysiological data--a practical approach},
  author={Widmann, Andreas and Schr{\"o}ger, Erich and Maess, Burkhard},
  journal={Journal of neuroscience methods},
  volume={250},
  pages={34--46},
  year={2015},
  publisher={Elsevier}
}

@article{eslami2024mitigate,
  title={Mitigate the gap: Investigating approaches for improving cross-modal alignment in clip},
  author={Eslami, Sedigheh and de Melo, Gerard},
  journal={arXiv preprint arXiv:2406.17639},
  year={2024}
}

@article{li2025tbme,
  title={Deep Learning for EEG-Based Visual Classification and Reconstruction: Panorama, Trends, Challenges and Opportunities},
  author={Li, Wei and Zhao, Penglu and Xu, Cheng and Hou, Yingting and Jiang, Wenhao and Song, Aiguo},
  journal={IEEE Transactions on Biomedical Engineering},
  volume={72},
  number={11},
  pages={3374--3390},
  year={2025}
}

@inproceedings{kim2025seeeeg,
  title={Seeeeg: Semantic-aware eeg-based multi-modal retrieval-augmented generation for high-fidelity visual brain decoding},
  author={Kim, Jun-Mo and Choi, Woohyeok and Park, Sang-Jun and Heo, Keun-Soo and Son, Young-Han and Oh, Ji-Hye and Shin, Dong-Hee and Kam, Tae-Eui},
  booktitle={Proceedings of the IEEE/CVF International Conference on Computer Vision},
  pages={4824--4833},
  year={2025}
}

@article{xu2026mindsae,
  title={MindSAE: Advancing semantic perception for M/EEG-based visual decoding via unified multimodal alignment framework},
  author={Xu, Chengjian and Song, Yonghao and Wang, Qiong and Zheng, Qingqing},
  journal={Biomedical Signal Processing and Control},
  volume={123},
  pages={110390},
  year={2026},
  publisher={Elsevier}
}

@inproceedings{zhang2026neurobridge,
  title={Neurobridge: Bio-inspired self-supervised eeg-to-image decoding via cognitive priors and bidirectional semantic alignment},
  author={Zhang, Wenjiang and Wang, Sifeng and Su, Yuwei and Li, Xinyu and Zhang, Chen and Zhong, Suyu},
  booktitle={Proceedings of the AAAI Conference on Artificial Intelligence},
  volume={40},
  number={21},
  pages={18028--18036},
  year={2026}
}

@article{xiong2025e2ivae,
  title={Interpretable Cross-Modal Alignment Network for EEG Visual Decoding With Algorithm Unrolling},
  author={Xiong, Daowen and Hu, Liangliang and Jin, Jiahao and Ding, Yikang and Tan, Congming and Zhang, Jing and Tian, Yin},
  journal={IEEE Transactions on Neural Networks and Learning Systems},
  volume={36},
  number={11},
  pages={19894--19908},
  year={2025}
}

@ARTICLE{ma2026neurodecoder,
  author={Ma, Wenxuan and Zhang, Hongxin and Li, Yexuan and Wei, Mingyi},
  journal={IEEE Journal of Biomedical and Health Informatics}, 
  title={NeuroDecoder: A new framework for image decoding and reconstruction of EEG signals}, 
  year={2026},
  volume={},
  number={},
  pages={1-14},
  doi={10.1109/JBHI.2026.3686624}}

@article{xiang2026regional,
  title={EEG-driven natural image reconstruction with regional semantic awareness},
  author={Xiang, Xin and Zhou, Wenhui and Zhu, Haonan and Li, Yunrui and Dai, Guojun and Lin, Lili},
  journal={Pattern Recognition},
  volume={172},
  pages={112589},
  year={2026}
}

@inproceedings{guo2025neuro3d,
  title={Neuro-3d: Towards 3d visual decoding from eeg signals},
  author={Guo, Zhanqiang and Wu, Jiamin and Song, Yonghao and Bu, Jiahui and Mai, Weijian and Zheng, Qihao and Ouyang, Wanli and Song, Chunfeng},
  booktitle={Proceedings of the Computer Vision and Pattern Recognition Conference},
  pages={23870--23880},
  year={2025}
}

@article{huang2025need,
  title={Need: Cross-subject and cross-task generalization for video and image reconstruction from eeg signals},
  author={Huang, Shuai and Luo, Huan and Jing, Haodong and Zhang, Qixian and Chang, Litao and Feng, Yating and Lin, Xiao and Qin, Chendong and Chen, Han and Jia, Shuwen and others},
  journal={Advances in Neural Information Processing Systems},
  volume={38},
  pages={173134--173173},
  year={2026}
}

@article{willett2021high,
  title={High-performance brain-to-text communication via handwriting},
  author={Willett, Francis R and Avansino, Donald T and Hochberg, Leigh R and Henderson, Jaimie M and Shenoy, Krishna V},
  journal={Nature},
  volume={593},
  number={7858},
  pages={249--254},
  year={2021},
  publisher={Nature Publishing Group UK London}
}

@article{gibson2022eeg,
  title={EEG variability: Task-driven or subject-driven signal of interest?},
  author={Gibson, Erin and Lobaugh, Nancy J and Joordens, Steve and McIntosh, Anthony R},
  journal={NeuroImage},
  volume={252},
  pages={119034},
  year={2022},
  publisher={Elsevier}
}

@article{xu2020crossdataset,
  title={Cross-dataset variability problem in EEG decoding with deep learning},
  author={Xu, Lichao and Xu, Minpeng and Ke, Yufeng and An, Xingwei and Liu, Shuang and Ming, Dong},
  journal={Frontiers in human neuroscience},
  volume={14},
  pages={103},
  year={2020},
  publisher={Frontiers Media SA}
}

@article{scotti2024reconstructing,
  title={Reconstructing the mind's eye: fmri-to-image with contrastive learning and diffusion priors},
  author={Scotti, Paul and Banerjee, Atmadeep and Goode, Jimmie and Shabalin, Stepan and Nguyen, Alex and Dempster, Aidan and Verlinde, Nathalie and Yundler, Elad and Weisberg, David and Norman, Kenneth and others},
  journal={Advances in Neural Information Processing Systems},
  volume={36},
  pages={24705--24728},
  year={2023}
}

@inproceedings{takagi2023high,
  title={High-resolution image reconstruction with latent diffusion models from human brain activity},
  author={Takagi, Yu and Nishimoto, Shinji},
  booktitle={Proceedings of the IEEE/CVF conference on computer vision and pattern recognition},
  pages={14453--14463},
  year={2023}
}

@article{grootswagers2019representational,
  title={The representational dynamics of visual objects in rapid serial visual processing streams},
  author={Grootswagers, Tijl and Robinson, Amanda K and Carlson, Thomas A},
  journal={NeuroImage},
  volume={188},
  pages={668--679},
  year={2019},
  publisher={Elsevier}
}

@inproceedings{wei2024mb2c,
  title={Mb2c: Multimodal bidirectional cycle consistency for learning robust visual neural representations},
  author={Wei, Yayun and Cao, Lei and Li, Hao and Dong, Yilin},
  booktitle={Proceedings of the 32nd ACM International Conference on Multimedia},
  pages={8992--9000},
  year={2024}
}

@ARTICLE{xiang2026latent3d,
  author={Xiang, Xin and Zhou, Wenhui and Dai, Guojun and Zhan, Shanggui},
  journal={IEEE Signal Processing Letters}, 
  title={Latent EEG-Vision Alignment for EEG-Driven 3D Object Reconstruction With Multi-View Stylistic Consistency}, 
  year={2026},
  volume={33},
  number={},
  pages={1906-1910},
  doi={10.1109/LSP.2026.3689379}}

\clearpage
\setcounter{equation}{0}
\setcounter{figure}{0}
\setcounter{table}{0}
\setcounter{page}{1}
\setcounter{section}{0}

\renewcommand{\theequation}{S\arabic{equation}}
\renewcommand{\thefigure}{S\arabic{figure}}
\renewcommand{\thetable}{S\arabic{table}}
\renewcommand{\thealgorithm}{S\arabic{algorithm}}
\renewcommand{\thesection}{S\Roman{section}}

\begin{center}
{\huge \textbf{Supplementary Material for ``STAMBRIDGE: Spectral-Temporal Amplitude-aware Mid-Feature Bridge for EEG Visual Decoding''}\par}
\vspace{1.5em}
{\large Jiahe Meng, Weiming Zeng*, Yueyang Li, Bo Chai, Hongjie Yan, Zhiguo Zhang, Wai Ting Siok, and Nizhuan Wang*\par}
\vspace{2.5em}
\end{center}

\begin{abstract}
This supplementary document provides comprehensive per-subject quantitative data and extensive qualitative visual reconstruction results to support the main manuscript.
\end{abstract}

\section{Inference Pipeline and Leakage Prevention}
To address concerns regarding potential target leakage during the generation-time inference, we provide the complete inference pseudocode in Algorithm \ref{alg:inference}. We strictly confirm that no ground-truth image or text features, nor the MFSB bridge features constructed during training, are available or utilized at test time. The generation relies exclusively on the raw EEG signal.

\begin{algorithm}[h]
\caption{Generation-Time Inference Pipeline of STAMBRIDGE}
\label{alg:inference}
\textbf{Input:} Raw EEG signals ($X_{eeg}$) from the test set \\
\textbf{Output:} Generated Image ($I_{gen}$) \\
\textbf{Note:} NO ground-truth image or text features are used.
\begin{algorithmic}[1]
\STATE \textit{\% Stage 1: Subject-Specific Adaptation \& Feature Extraction}
\STATE $E_{adapted} \leftarrow \text{Subject\_Specific\_Linear\_Layer}(X_{eeg})$
\STATE $E_{features} \leftarrow \text{STAM\_Module}(E_{adapted})$
\STATE \textit{\% Stage 2: Projection to Semantic Space}
\STATE \textit{\% Crucially, we bypass the MFSB module entirely.}
\STATE $Z_{eeg\_proj} \leftarrow \text{EEG\_Projector}(E_{features})$
\STATE \textit{\% Stage 3: Diffusion Prior Generation}
\STATE $Z_{image\_pred} \leftarrow \text{Diffusion\_Prior}(Z_{eeg\_proj})$
\STATE \textit{\% Stage 4: Visual Decoding}
\STATE $I_{gen} \leftarrow \text{SD\_Decoder}(Z_{image\_pred})$
\RETURN $I_{gen}$
\end{algorithmic}
\end{algorithm}

\section{Additional Quantitative Results}
In the main text, we provided the mean and standard deviation of the model's performance to highlight its robustness. Here, we present the comprehensive per-subject breakdowns. 

\textbf{Table S1} provides the detailed generative performance metrics (including both high-level semantic indicators and low-level structural metrics such as PixCorr and SSIM) across all 10 subjects.

\begin{table*}[!htbp]
\centering
\caption{Comprehensive quantitative comparison of generative performance across all 10 subjects. We re-implemented the ATM baseline using identical generative pipelines (50-step Diffusion Prior + 4-step IP-Adapter) to ensure fair comparison. Both structural/low-level metrics and high-level semantic metrics are reported.}
\label{tab:S1_comprehensive_generation}
\resizebox{\textwidth}{!}{
\begin{tabular}{llcccccccc}
\toprule
\textbf{Subject} & \textbf{Method} & \textbf{PixCorr $\uparrow$} & \textbf{SSIM $\uparrow$} & \textbf{AlexNet(2) $\uparrow$} & \textbf{AlexNet(5) $\uparrow$} & \textbf{Inception $\uparrow$} & \textbf{CLIP $\uparrow$} & \textbf{EffNet-B $\uparrow$} & \textbf{SwAV $\downarrow$} \\
\midrule
\multirow{2}{*}{Sub-01} 
 & ATM & \textbf{0.125} & 0.323 & 0.775 & \textbf{0.864} & \textbf{0.707} & 0.724 & 0.883 & \textbf{0.513} \\
 & STAMBRIDGE & 0.119 & \textbf{0.335} & \textbf{0.789} & 0.863 & 0.698 & 0.724 & \textbf{0.897} & 0.526 \\
\midrule
\multirow{2}{*}{Sub-02} 
 & ATM & 0.130 & \textbf{0.337} & 0.792 & \textbf{0.853} & 0.674 & 0.693 & \textbf{0.918} & \textbf{0.526} \\
 & STAMBRIDGE & \textbf{0.146} & 0.321 & \textbf{0.803} & 0.852 & \textbf{0.716} & \textbf{0.734} & 0.893 & 0.531 \\
\midrule
\multirow{2}{*}{Sub-03} 
 & ATM & 0.133 & \textbf{0.285} & 0.770 & \textbf{0.857} & 0.697 & 0.720 & \textbf{0.897} & 0.532 \\
 & STAMBRIDGE & \textbf{0.145} & 0.282 & \textbf{0.773} & \textbf{0.857} & \textbf{0.736} & \textbf{0.739} & 0.891 & \textbf{0.514} \\
\midrule
\multirow{2}{*}{Sub-04} 
 & ATM & 0.120 & 0.303 & \textbf{0.793} & 0.885 & 0.769 & 0.779 & 0.857 & 0.489 \\
 & STAMBRIDGE & \textbf{0.122} & \textbf{0.325} & 0.786 & \textbf{0.894} & \textbf{0.775} & \textbf{0.788} & \textbf{0.864} & \textbf{0.487} \\
\midrule
\multirow{2}{*}{Sub-05} 
 & ATM & 0.109 & 0.320 & \textbf{0.797} & \textbf{0.863} & \textbf{0.692} & \textbf{0.720} & 0.900 & \textbf{0.523} \\
 & STAMBRIDGE & \textbf{0.115} & \textbf{0.341} & 0.791 & 0.822 & 0.673 & 0.670 & \textbf{0.905} & 0.538 \\
\midrule
\multirow{2}{*}{Sub-06} 
 & ATM & 0.151 & \textbf{0.315} & 0.785 & 0.853 & 0.689 & 0.686 & \textbf{0.905} & \textbf{0.530} \\
 & STAMBRIDGE & \textbf{0.160} & 0.289 & \textbf{0.801} & \textbf{0.861} & \textbf{0.697} & \textbf{0.702} & 0.903 & 0.545 \\
\midrule
\multirow{2}{*}{Sub-07} 
 & ATM & 0.098 & \textbf{0.315} & 0.767 & 0.852 & 0.660 & 0.707 & \textbf{0.900} & 0.530 \\
 & STAMBRIDGE & \textbf{0.117} & 0.283 & 0.767 & \textbf{0.869} & \textbf{0.694} & \textbf{0.747} & 0.882 & \textbf{0.524} \\
\midrule
\multirow{2}{*}{Sub-08} 
 & ATM & \textbf{0.166} & \textbf{0.318} & 0.792 & 0.874 & 0.717 & 0.734 & \textbf{0.886} & 0.523 \\
 & STAMBRIDGE & 0.159 & 0.304 & \textbf{0.802} & \textbf{0.879} & \textbf{0.765} & \textbf{0.759} & 0.861 & \textbf{0.502} \\
\midrule
\multirow{2}{*}{Sub-09} 
 & ATM & 0.114 & \textbf{0.311} & \textbf{0.758} & \textbf{0.838} & \textbf{0.712} & \textbf{0.679} & \textbf{0.904} & \textbf{0.516} \\
 & STAMBRIDGE & \textbf{0.121} & 0.281 & 0.732 & 0.825 & 0.673 & 0.674 & 0.897 & 0.548 \\
\midrule
\multirow{2}{*}{Sub-10} 
 & ATM & \textbf{0.149} & \textbf{0.366} & 0.797 & 0.862 & 0.718 & 0.746 & \textbf{0.871} & 0.482 \\
 & STAMBRIDGE & 0.131 & 0.345 & \textbf{0.798} & \textbf{0.876} & \textbf{0.741} & \textbf{0.763} & 0.871 & \textbf{0.477} \\
\midrule
\multicolumn{10}{c}{\textbf{Overall Mean Performance (10 Subjects)}} \\
\midrule
\multirow{2}{*}{Average}
 & ATM & 0.130 & \textbf{0.320} & 0.783 & \textbf{0.860} & 0.704 & 0.719 & \textbf{0.892} & \textbf{0.516} \\
 & STAMBRIDGE & \textbf{0.134} & 0.311 & \textbf{0.784} & \textbf{0.860} & \textbf{0.717} & \textbf{0.730} & 0.886 & 0.519 \\
\bottomrule
\end{tabular}
}
\end{table*}

\begin{table*}[!t] 
\centering
\caption{Comprehensive Statistical Significance and Effect Size Analysis (Paired t-test, $N=10$).}
\label{tab:supp_statistics}
\begin{tabular}{llcccc}
\toprule
\textbf{Experimental Group} & \textbf{Metric} & \textbf{Absolute Gain (\%)} & \textbf{$p$-value} & \textbf{Cohen's $d$} & \textbf{Significance} \\
\midrule
\multicolumn{6}{c}{\textbf{Part I: Main Model Comparison}} \\
\midrule
\textbf{Neural-MCRL (Baseline) vs. STAMBRIDGE} & \textbf{Top-1} & \textbf{+3.45} & \textbf{0.00057} & \textbf{1.64} & \textbf{***} \\
& \textbf{Top-5} & \textbf{+6.25} & \textbf{0.00002} & \textbf{2.57} & \textbf{***} \\

\midrule
\multicolumn{6}{c}{\textbf{Part II: Cross-model Generalization of MFSB (Plugin)}} \\
\midrule
\multirow{2}{*}{NICE vs. NICE + MFSB} & Top-1 & +1.00 & 0.12523 & 0.53 & n.s. \\
& Top-5 & +3.25 & 0.00041 & 1.72 & *** \\
\multirow{2}{*}{ATM vs. ATM + MFSB} & Top-1 & +1.20 & 0.27423 & 0.37 & n.s. \\
& Top-5 & +3.00 & 0.03435 & 0.79 & * \\
\multirow{2}{*}{Neural-MCRL vs. Neural-MCRL + MFSB} & Top-1 & +1.70 & 0.01050 & 1.02 & * \\
& Top-5 & +3.00 & 0.02338 & 0.86 & * \\

\midrule
\multicolumn{6}{c}{\textbf{Part III: Component Ablation Study}} \\
\midrule
\multirow{2}{*}{Baseline vs. +STAM Only} & Top-1 & +2.90 & 0.00084 & 1.55 & *** \\
& Top-5 & +3.90 & 0.00534 & 1.15 & ** \\
\multirow{2}{*}{Baseline vs. +Bridge Only} & Top-1 & +1.55 & 0.07922 & 0.63 & n.s. \\
& Top-5 & +4.15 & 0.01307 & 0.98 & * \\
\midrule
\multirow{2}{*}{+STAM Only vs. Full Model (STAMBRIDGE)} & Top-1 & +0.55 & 0.51825 & 0.21 & n.s. \\
& Top-5 & +2.35 & 0.01897 & 0.90 & * \\
\bottomrule
\multicolumn{6}{l}{\scriptsize{n.s.: Not Significant ($p \ge 0.05$); $^{*}p < 0.05$; $^{**}p < 0.01$; $^{***}p < 0.001$}} \\
\multicolumn{6}{l}{\scriptsize{Cohen's $d$ is interpreted following conventional thresholds: 0.2 (small), 0.5 (medium), and 0.8 (large).}} \\
\end{tabular}
\end{table*}

\begin{table*}[!t]
\centering
\caption{Overall 200-way zero-shot retrieval accuracy (\%) on the THINGS-EEG dataset (Subject-Dependent). Results are reported as mean $\pm$ std across 10 subjects. Statistical significance compared to the strongest baseline (Neural-MCRL) is denoted by asterisks ($^{***}p < 0.001$). Detailed statistical testing, including exact $p$-values and Cohen's $d$ effect sizes, is provided in Supplementary Table S2.}
\label{tab:S3_retrieval_per_subject}
\resizebox{\textwidth}{!}{
\begin{tabular}{lcccccccccccccccccccccc}
\toprule
\multirow{2}{*}{Method} & \multicolumn{2}{c}{S1} & \multicolumn{2}{c}{S2} & \multicolumn{2}{c}{S3} & \multicolumn{2}{c}{S4} & \multicolumn{2}{c}{S5} & \multicolumn{2}{c}{S6} & \multicolumn{2}{c}{S7} & \multicolumn{2}{c}{S8} & \multicolumn{2}{c}{S9} & \multicolumn{2}{c}{S10} & \multicolumn{2}{c}{\textbf{\mstd{Mean}{Std}}} \\
\cmidrule(lr){2-3} \cmidrule(lr){4-5} \cmidrule(lr){6-7} \cmidrule(lr){8-9} \cmidrule(lr){10-11} \cmidrule(lr){12-13} \cmidrule(lr){14-15} \cmidrule(lr){16-17} \cmidrule(lr){18-19} \cmidrule(lr){20-21} \cmidrule(lr){22-23}
 & T1 & T5 & T1 & T5 & T1 & T5 & T1 & T5 & T1 & T5 & T1 & T5 & T1 & T5 & T1 & T5 & T1 & T5 & T1 & T5 & \textbf{T1} & \textbf{T5} \\
\midrule
BraVL        & 6.1  & 17.9 & 4.9  & 14.9 & 5.6  & 17.4 & 4.9  & 15.1 & 4.0  & 13.4 & 6.0  & 18.1 & 6.5  & 20.3 & 8.8  & 23.7 & 4.3  & 13.9 & 7.0  & 13.7 & \mstd{5.81}{1.42}  & \mstd{16.84}{3.32} \\
NICE         & 11.5 & 32.0 & 12.5 & 35.5 & 12.5 & 41.5 & 18.5 & 48.0 & 9.5  & 28.5 & 15.0 & 42.5 & 16.0 & 41.0 & 19.5 & 51.0 & 16.5 & 37.5 & 16.5 & 43.5 & \mstd{14.80}{3.21} & \mstd{40.10}{6.91} \\
MB2C         & 23.0 & 56.0 & 23.0 & 56.0 & 30.0 & 61.5 & 21.5 & 48.5 & 21.5 & 48.5 & 32.0 & 61.5 & 28.5 & 59.5 & 40.5 & 70.0 & 27.5 & 59.0 & 29.5 & 69.0 & \mstd{27.70}{5.89} & \mstd{58.95}{7.24} \\
ATM          & 23.0 & 48.0 & 19.5 & 42.5 & 33.0 & 58.5 & 29.5 & 57.0 & 26.5 & 46.5 & 25.5 & 61.0 & 28.5 & 59.5 & 40.5 & 71.0 & 30.0 & 53.0 & 37.5 & 66.5 & \mstd{29.35}{6.37} & \mstd{56.35}{8.97} \\
Neural-MCRL  & 28.5 & 56.5 & 25.5 & 56.5 & 30.0 & 59.0 & 34.0 & 62.0 & 24.0 & 51.0 & 27.0 & 56.5 & 30.5 & 58.5 & 42.5 & 76.5 & 30.5 & 53.5 & 38.0 & 67.0 & \mstd{31.05}{5.71} & \mstd{59.70}{7.36} \\
\midrule
\textbf{STAMBRIDGE (Ours)} & \textbf{33.0} & \textbf{67.5} & \textbf{31.5} & \textbf{64.0} & \textbf{36.0} & \textbf{66.5} & \textbf{37.0} & \textbf{67.5} & \textbf{25.5} & \textbf{58.0} & \textbf{29.5} & \textbf{62.5} & \textbf{33.5} & \textbf{63.0} & \textbf{48.5} & \textbf{79.0} & \textbf{32.5} & \textbf{61.0} & \textbf{38.0} & \textbf{70.5} & \textbf{\mstd{34.50}{6.14}$^{***}$} & \textbf{\mstd{65.95}{5.85}$^{***}$} \\
\bottomrule
\end{tabular}
}
\end{table*}

\begin{table*}[!htbp]
\centering
\caption{Comprehensive per-subject zero-shot retrieval accuracy (\%) for additional ablation studies and baseline comparisons. This table details the performance distributions that correspond to the ablation summaries provided in the main manuscript.}
\label{tab:S4_comprehensive_ablation}
\resizebox{\textwidth}{!}{
\begin{tabular}{lcccccccccccccccccccccc}
\toprule
\multirow{2}{*}{Method} & \multicolumn{2}{c}{S1} & \multicolumn{2}{c}{S2} & \multicolumn{2}{c}{S3} & \multicolumn{2}{c}{S4} & \multicolumn{2}{c}{S5} & \multicolumn{2}{c}{S6} & \multicolumn{2}{c}{S7} & \multicolumn{2}{c}{S8} & \multicolumn{2}{c}{S9} & \multicolumn{2}{c}{S10} & \multicolumn{2}{c}{\textbf{\mstd{Mean}{Std}}} \\
\cmidrule(lr){2-3} \cmidrule(lr){4-5} \cmidrule(lr){6-7} \cmidrule(lr){8-9} \cmidrule(lr){10-11} \cmidrule(lr){12-13} \cmidrule(lr){14-15} \cmidrule(lr){16-17} \cmidrule(lr){18-19} \cmidrule(lr){20-21} \cmidrule(lr){22-23}
 & T1 & T5 & T1 & T5 & T1 & T5 & T1 & T5 & T1 & T5 & T1 & T5 & T1 & T5 & T1 & T5 & T1 & T5 & T1 & T5 & \textbf{T1} & \textbf{T5} \\
\midrule
\multicolumn{23}{l}{\textit{(a) Training Strategy Optimization}} \\
Truly Sequential & 31.5 & 66.0 & 28.5 & 56.0 & 37.5 & 69.5 & 33.5 & 68.0 & 25.0 & 53.0 & 31.0 & 65.5 & 32.0 & 65.0 & 42.5 & 75.5 & 33.5 & 59.0 & 37.5 & 69.5 & \mstd{33.25}{4.97} & \mstd{64.70}{6.84} \\
Simultaneous Joint & 31.5 & 68.0 & 28.5 & 60.5 & 32.5 & 63.5 & 38.0 & 72.5 & 20.5 & 52.0 & 35.5 & 69.0 & 35.0 & 68.5 & 45.0 & 78.5 & 33.5 & 58.5 & 36.0 & 71.0 & \mstd{33.60}{6.36} & \mstd{66.20}{7.68} \\
\midrule
\multicolumn{23}{l}{\textit{(b) MFSB Visual Drop Rate (Shortcut Control)}} \\
Drop = 1.0 (Pure EEG) & 29.0 & 58.5 & 31.0 & 59.0 & 36.0 & 66.0 & 30.0 & 57.5 & 22.5 & 51.5 & 28.0 & 58.5 & 37.0 & 65.0 & 47.5 & 79.0 & 28.0 & 55.0 & 37.5 & 70.5 & \mstd{32.65}{7.02} & \mstd{62.05}{8.16} \\
Drop = 0.8 & 32.0 & 62.5 & 27.0 & 56.0 & 36.5 & 67.0 & 38.0 & 68.5 & 19.5 & 50.0 & 30.0 & 63.5 & 34.0 & 64.5 & 45.5 & 79.0 & 31.5 & 63.0 & 35.5 & 70.5 & \mstd{32.95}{6.92} & \mstd{64.45}{7.88} \\
Drop = 0.6 & 35.0 & 65.0 & 27.0 & 60.5 & 34.0 & 66.5 & 36.5 & 70.5 & 18.5 & 44.5 & 31.0 & 59.0 & 33.0 & 65.5 & 47.0 & 77.0 & 35.0 & 63.5 & 36.0 & 70.5 & \mstd{33.30}{7.27} & \mstd{64.25}{8.69} \\
\midrule
\multicolumn{23}{l}{\textit{(c) Cross-Modal Fusion Mechanisms}} \\
Standard Cross-Attn & 33.0 & 62.0 & 29.5 & 67.5 & 36.0 & 68.0 & 36.0 & 72.0 & 23.0 & 52.5 & 33.5 & 63.5 & 33.0 & 63.0 & 42.5 & 77.0 & 30.0 & 58.5 & 40.0 & 73.5 & \mstd{33.65}{5.52} & \mstd{65.75}{7.37} \\
Simple Weighted Avg & 33.0 & 63.0 & 28.5 & 61.0 & 36.0 & 64.5 & 40.5 & 69.0 & 21.0 & 51.0 & 32.5 & 63.0 & 35.0 & 68.5 & 47.0 & 81.0 & 26.5 & 60.5 & 37.0 & 69.5 & \mstd{33.70}{7.33} & \mstd{65.10}{7.78} \\
Concatenation Fusion & 30.0 & 63.5 & 28.0 & 59.5 & 37.0 & 73.0 & 39.0 & 68.5 & 21.0 & 50.5 & 34.0 & 68.0 & 35.0 & 68.0 & 46.0 & 76.5 & 31.0 & 64.0 & 40.5 & 70.5 & \mstd{34.15}{7.08} & \mstd{66.20}{7.35} \\
\midrule
\multicolumn{23}{l}{\textit{(d) Spectral-Temporal Modulation Mechanisms}} \\
Hard Spectral Masking & 34.0 & 66.5 & 26.0 & 53.5 & 38.0 & 72.0 & 38.0 & 67.5 & 23.5 & 58.5 & 30.5 & 63.5 & 34.5 & 67.5 & 44.0 & 76.5 & 27.5 & 62.0 & 38.5 & 69.0 & \mstd{33.45}{6.48} & \mstd{65.65}{6.61} \\
Tapered Filtering & 31.0 & 63.5 & 27.5 & 60.5 & 37.5 & 68.0 & 38.0 & 69.5 & 27.0 & 56.5 & 28.5 & 64.0 & 34.5 & 68.5 & 43.5 & 78.0 & 33.0 & 64.5 & 38.5 & 73.5 & \mstd{33.90}{5.48} & \mstd{66.65}{6.25} \\
Standard Channel Attn & 27.5 & 62.5 & 29.0 & 61.5 & 35.5 & 68.5 & 38.0 & 68.0 & 24.5 & 53.0 & 30.5 & 65.0 & 33.5 & 67.5 & 43.0 & 78.0 & 33.5 & 59.5 & 40.0 & 71.0 & \mstd{33.50}{5.80} & \mstd{65.45}{6.86} \\
Time-Domain Gating & 33.5 & 64.5 & 32.0 & 66.5 & 36.0 & 64.0 & 36.0 & 71.0 & 25.5 & 57.0 & 33.0 & 69.5 & 34.5 & 64.0 & 44.0 & 75.0 & 30.0 & 64.0 & 36.5 & 69.0 & \mstd{34.10}{4.81} & \mstd{66.45}{4.95} \\
\midrule
\textbf{STAMBRIDGE (Ours)} & 33.0 & 67.5 & 31.5 & 64.0 & 36.0 & 66.5 & 37.0 & 67.5 & 25.5 & 58.0 & 29.5 & 62.5 & 33.5 & 63.0 & 48.5 & 79.0 & 32.5 & 61.0 & 38.0 & 70.5 & \textbf{\mstd{34.50}{6.14}} & \textbf{\mstd{65.95}{5.85}} \\
\bottomrule
\end{tabular}
}
\end{table*}

\clearpage
\section{Additional Qualitative Results}
This section provides extensive visualizations to validate the structural integrity of the learned EEG embeddings and the efficacy of the generative prior.

\textbf{Figure S1} and \textbf{Figure S2} illustrate the Top-10 zero-shot image retrieval results and the t-SNE latent space alignment, respectively. 

Furthermore, to demonstrate the consistency of STAMBRIDGE across subjects, we provide the qualitative semantic image reconstruction results for all 10 subjects (\textbf{Figures S4 to S13}) under the rapid serial visual presentation (RSVP) paradigm. For each subject, the reconstructed images are organized into \textbf{Best}, \textbf{Median}, and \textbf{Worst} cases according to their high-level semantic similarity to the corresponding ground-truth visual stimuli.

\begin{figure}[!h]
  \centering
  \adjustbox{max width=\columnwidth}{%
    \includegraphics{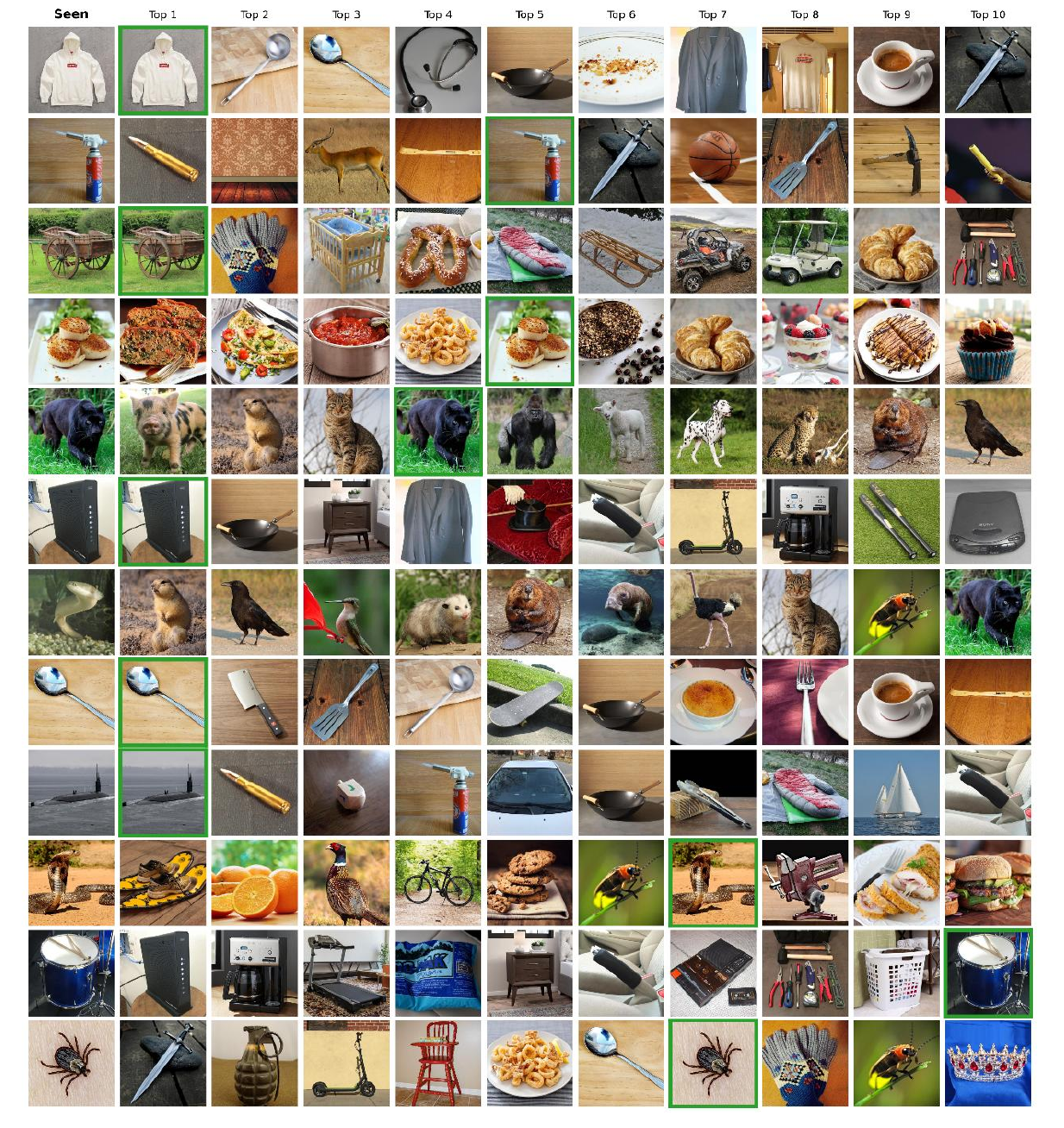}%
  }
  \caption{Top-10 zero-shot image retrieval results based on unseen EEG queries. The green boxes indicate correct matches (Top-1). Even for incorrect Top-1 retrievals, the subsequent candidates exhibit highly consistent semantics, demonstrating the robustness of our learned representations.}
  \label{fig:S1_retrieval}
\end{figure}

\begin{figure}[!h]
  \centering 
  \adjustbox{max width=\linewidth}{%
    \includegraphics{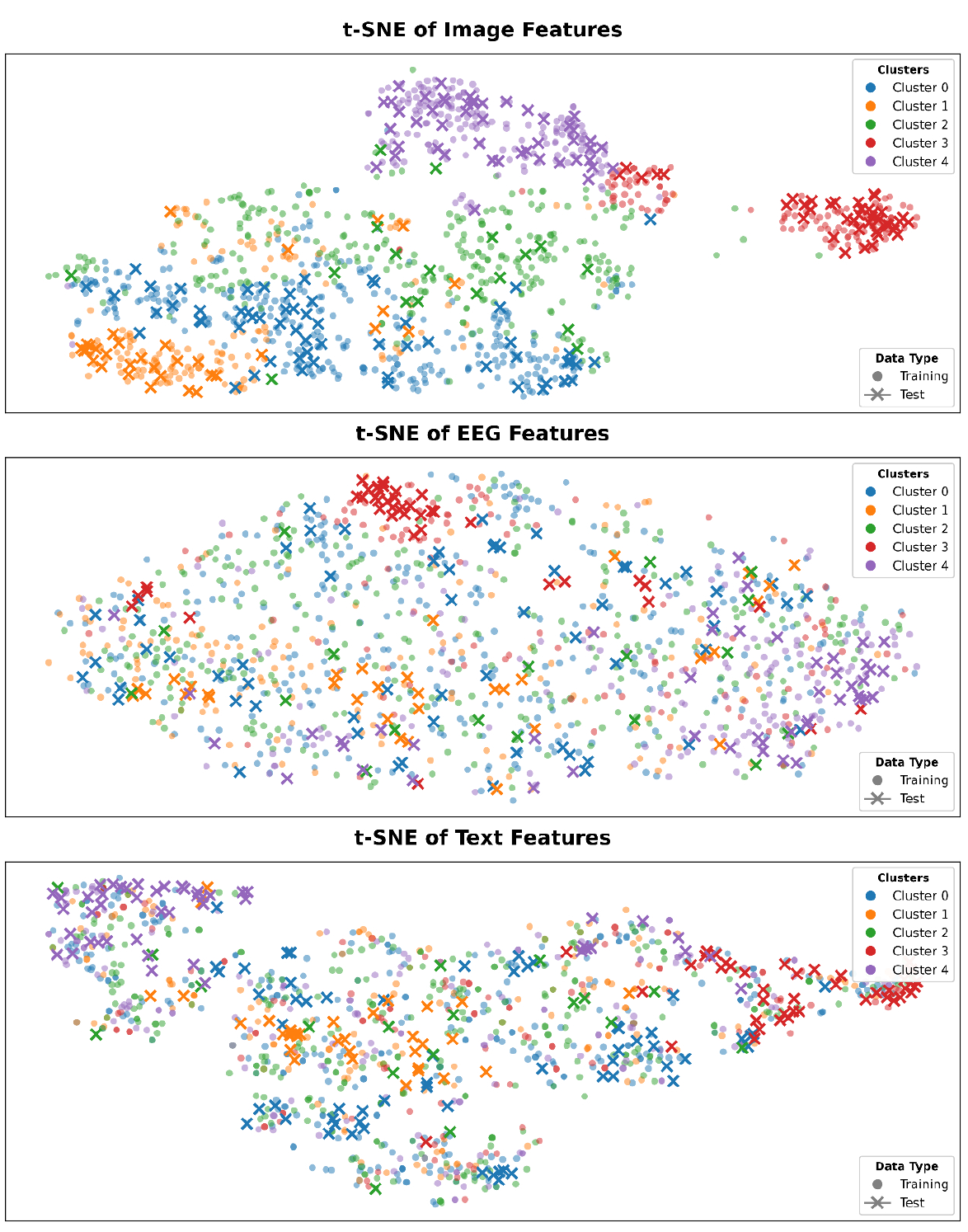}%
  }
  \caption{t-SNE visualization of Image, EEG, and Text features. STAMBRIDGE successfully aligns EEG features into semantically consistent clusters that correspond closely to the visual and textual manifolds.}
  \label{fig:S2_tsne}
\end{figure}

\begin{figure*}[!h]
  \centering 
  \adjustbox{max width=\linewidth}{%
    \includegraphics{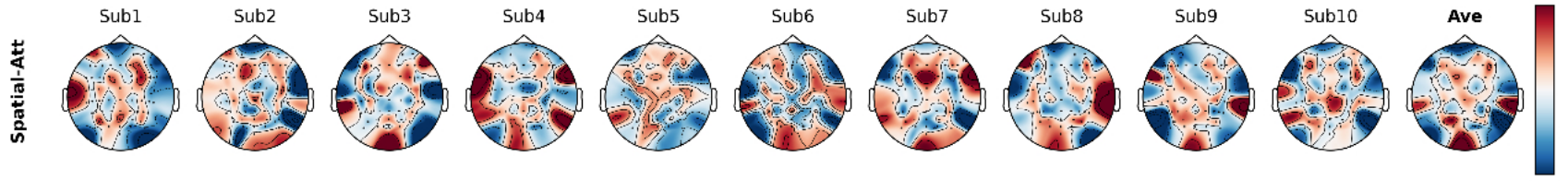}%
  }
  \caption{Visualization of the learned spatial attention maps across 10 subjects (Sub1--Sub10) and their grand average (Ave). The activation heatmaps consistently highlight the occipital and parietal regions, confirming that the model captures biologically meaningful visual processing signals rather than noise.}
  \label{fig:S3_spatial_att}
\end{figure*}

\clearpage

\begin{figure*}[h!]
  \centering
  \includegraphics[width=0.95\textwidth]{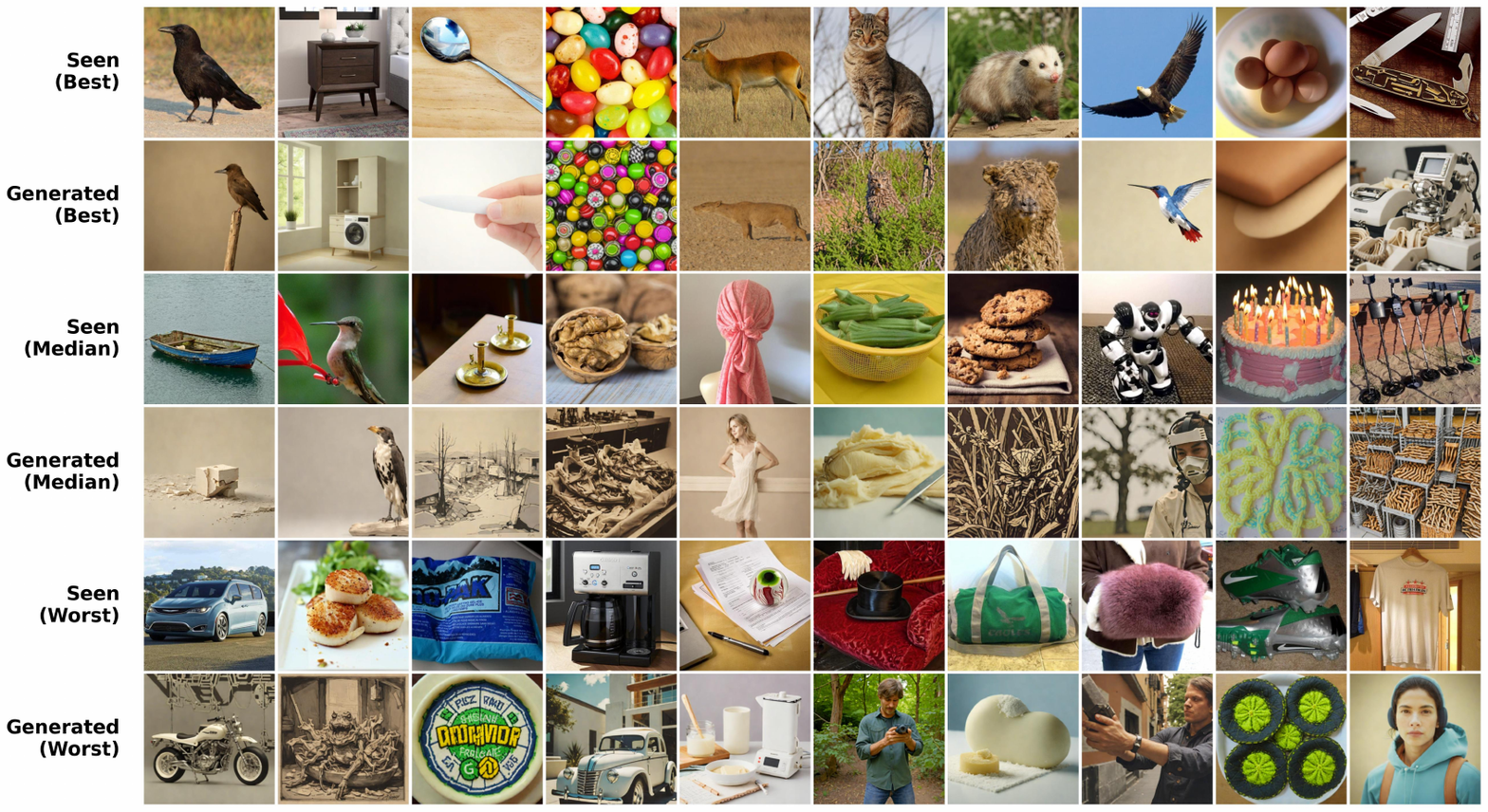}
  \caption{Qualitative results of semantic image reconstruction for \textbf{Subject 01}. The rows correspond to the Best, Median, and Worst reconstruction cases.}
  \label{fig:S3}
\end{figure*}

\begin{figure*}[h!]
  \centering
  \includegraphics[width=0.95\textwidth]{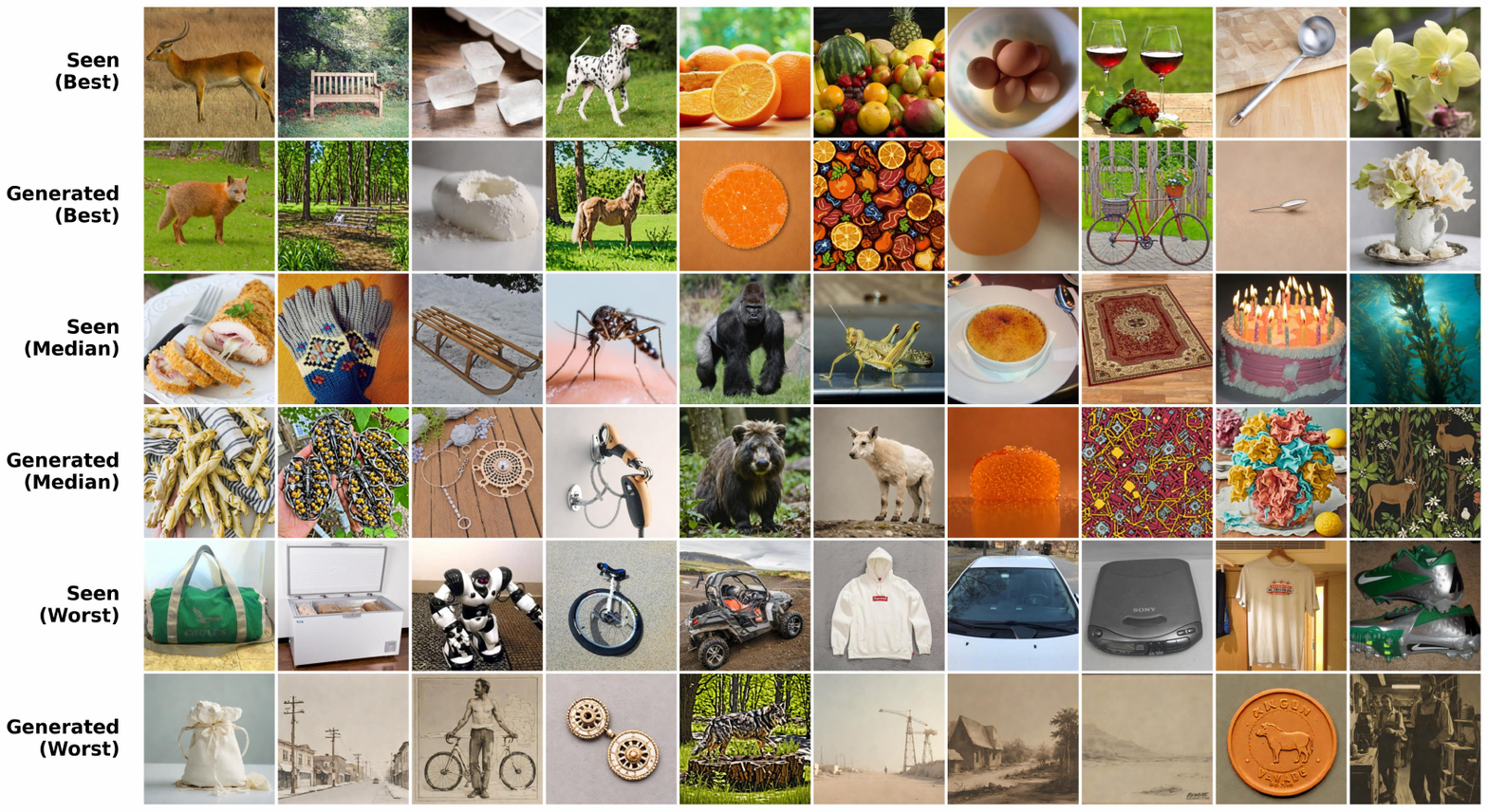}
  \caption{Qualitative results of semantic image reconstruction for \textbf{Subject 02}. The rows correspond to the Best, Median, and Worst reconstruction cases.}
  \label{fig:S4}
\end{figure*}

\begin{figure*}[h!]
  \centering
  \includegraphics[width=0.95\textwidth]{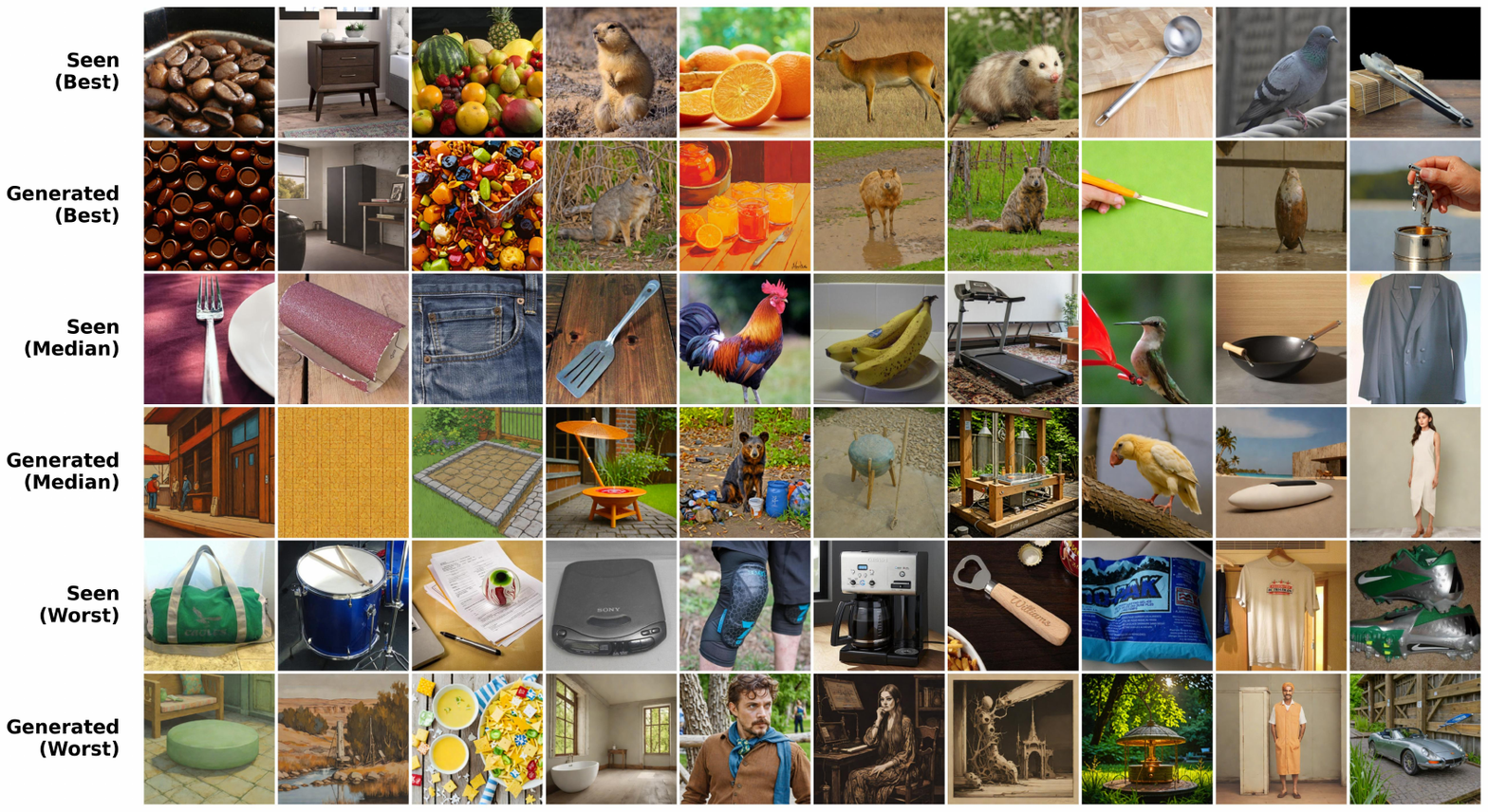}
  \caption{Qualitative results of semantic image reconstruction for \textbf{Subject 03}. The rows correspond to the Best, Median, and Worst reconstruction cases.}
  \label{fig:S5}
\end{figure*}

\begin{figure*}[h!]
  \centering
  \includegraphics[width=0.95\textwidth]{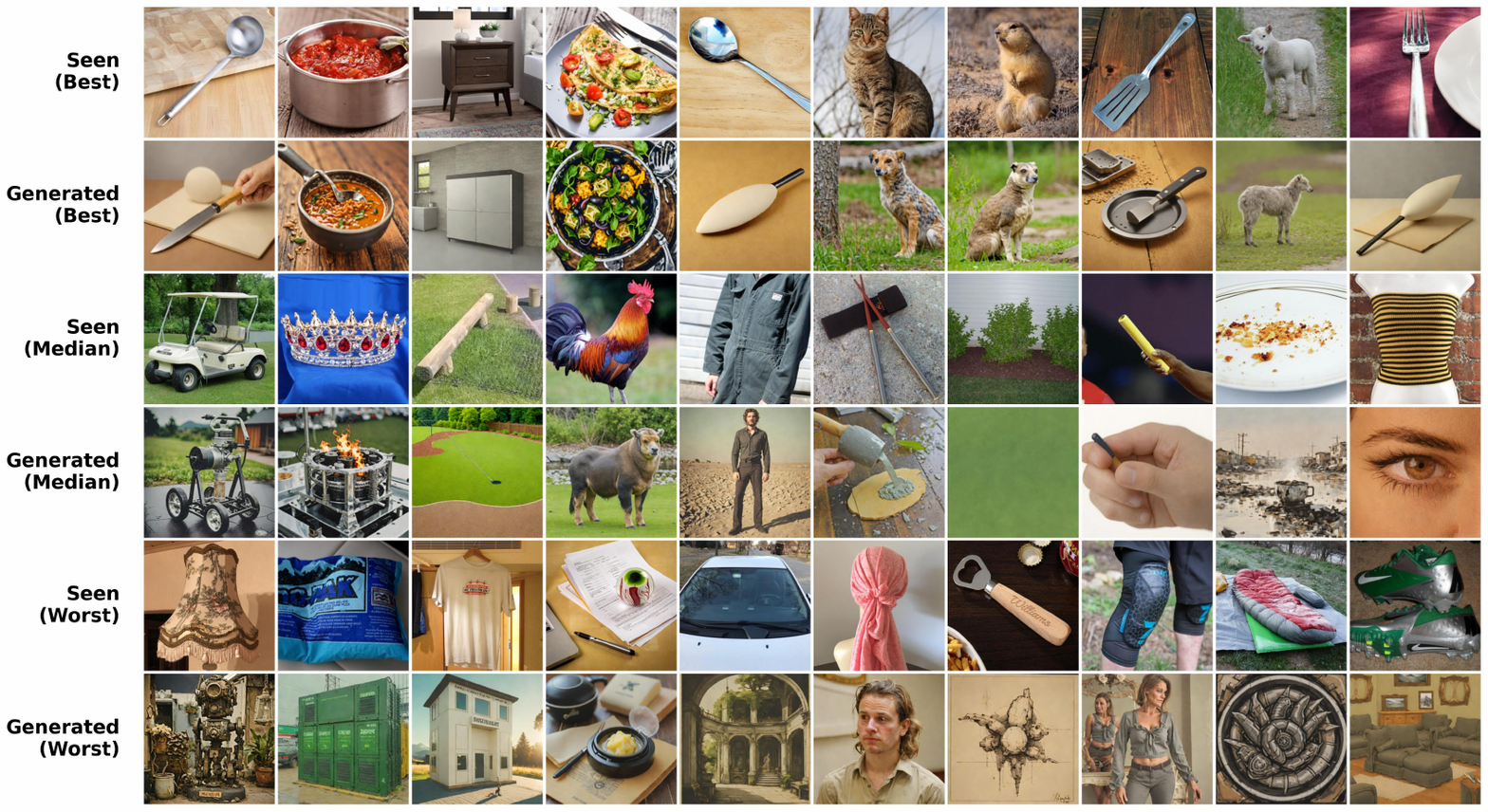}
  \caption{Qualitative results of semantic image reconstruction for \textbf{Subject 04}. The rows correspond to the Best, Median, and Worst reconstruction cases.}
  \label{fig:S6}
\end{figure*}

\begin{figure*}[h!]
  \centering
  \includegraphics[width=0.95\textwidth]{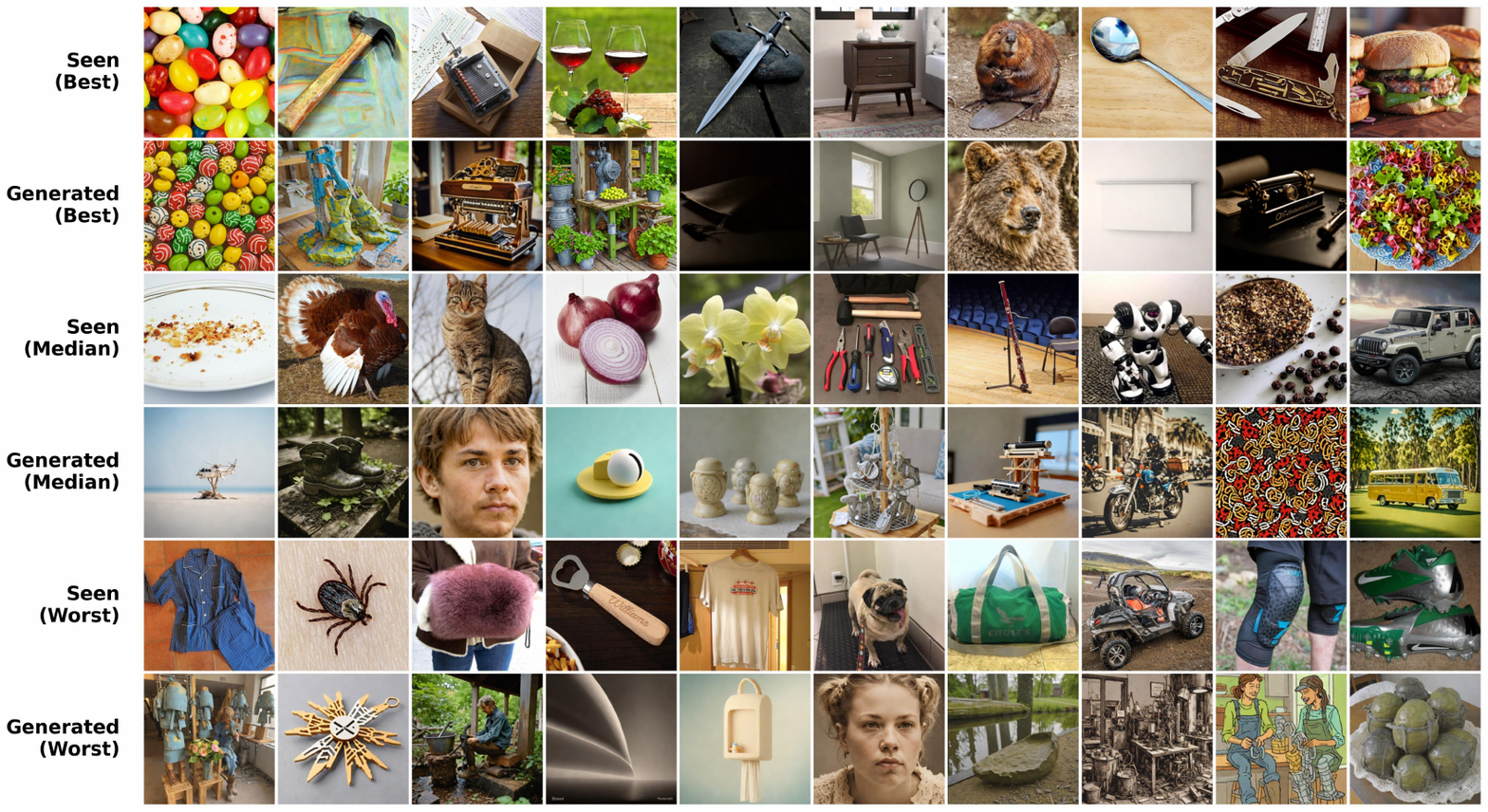}
  \caption{Qualitative results of semantic image reconstruction for \textbf{Subject 05}. The rows correspond to the Best, Median, and Worst reconstruction cases.}
  \label{fig:S7}
\end{figure*}

\begin{figure*}[h!]
  \centering
  \includegraphics[width=0.95\textwidth]{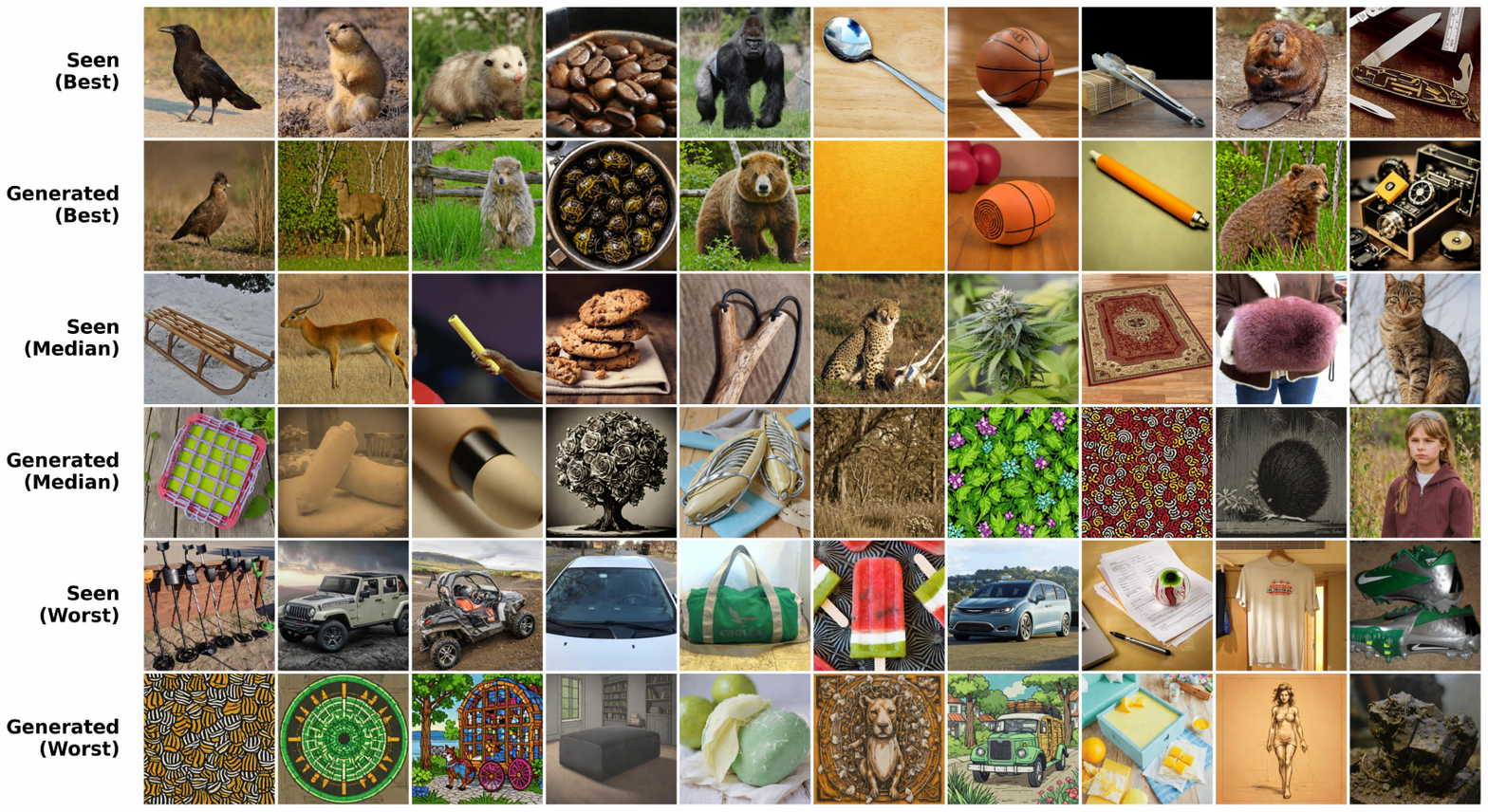}
  \caption{Qualitative results of semantic image reconstruction for \textbf{Subject 06}. The rows correspond to the Best, Median, and Worst reconstruction cases.}
  \label{fig:S8}
\end{figure*}

\begin{figure*}[h!]
  \centering
  \includegraphics[width=0.95\textwidth]{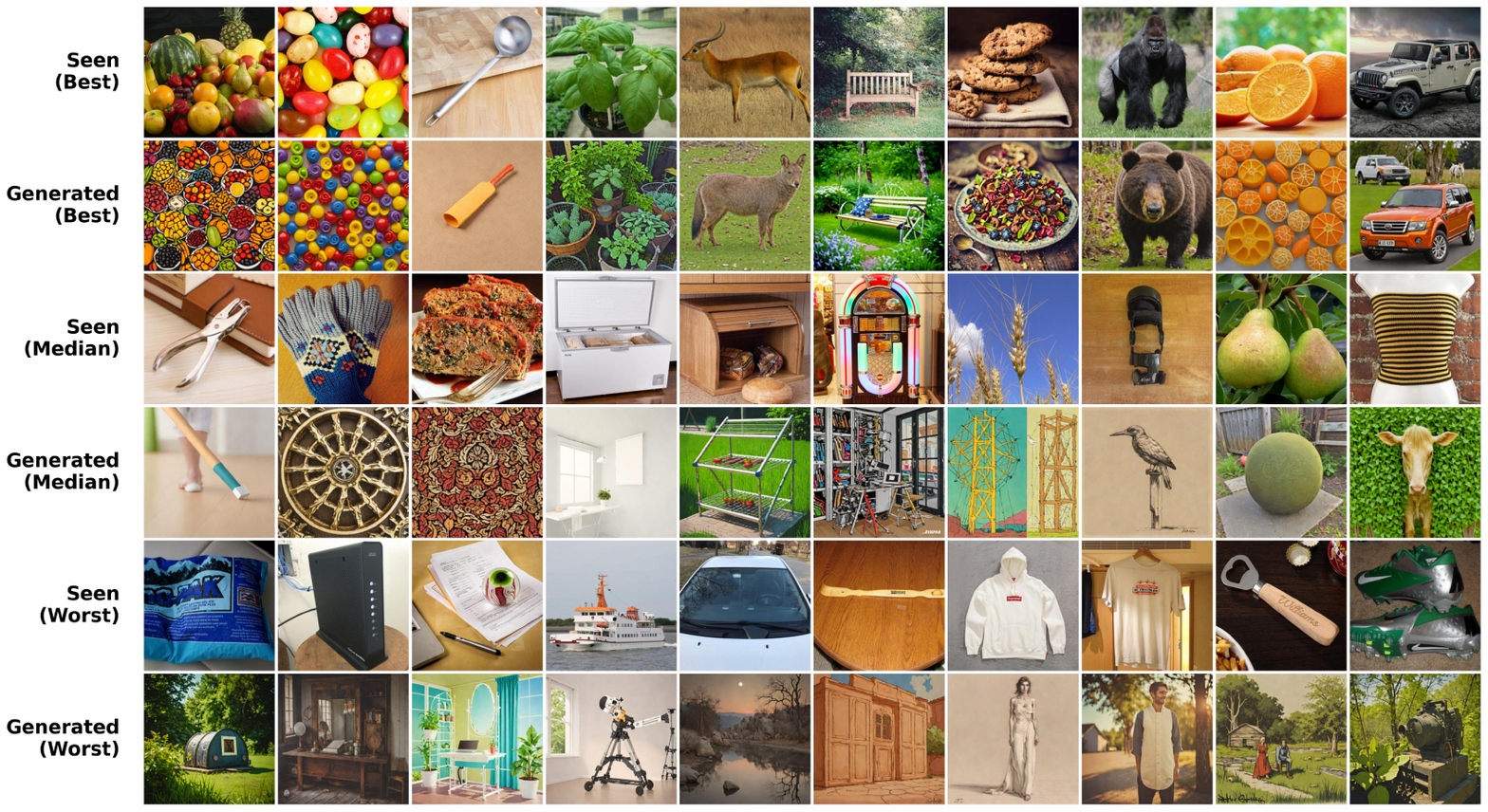}
  \caption{Qualitative results of semantic image reconstruction for \textbf{Subject 07}. The rows correspond to the Best, Median, and Worst reconstruction cases.}
  \label{fig:S9}
\end{figure*}

\begin{figure*}[h!]
  \centering
  \includegraphics[width=0.95\textwidth]{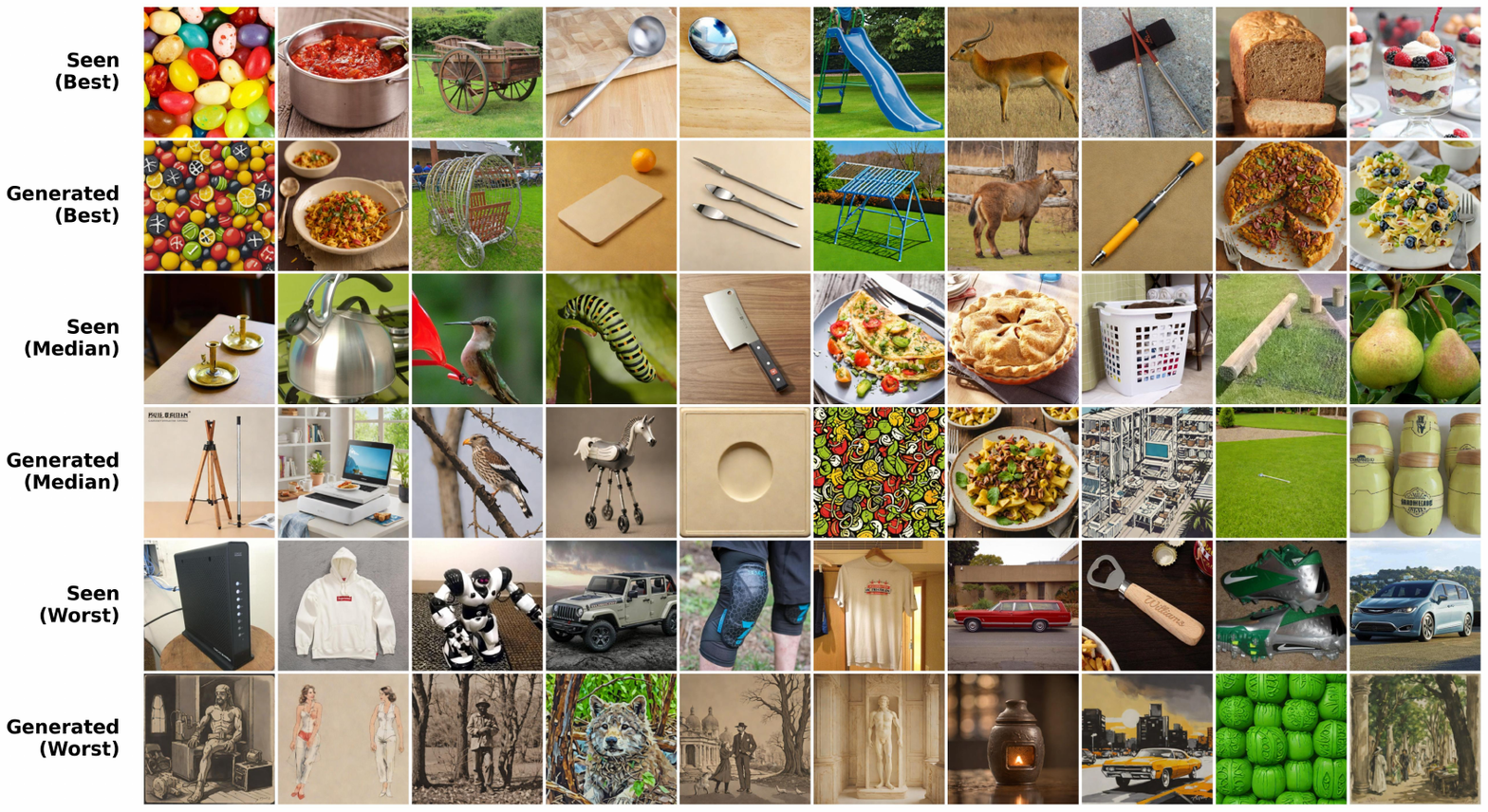}
  \caption{Qualitative results of semantic image reconstruction for \textbf{Subject 08}. The rows correspond to the Best, Median, and Worst reconstruction cases.}
  \label{fig:S10}
\end{figure*}

\begin{figure*}[h!]
  \centering
  \includegraphics[width=0.95\textwidth]{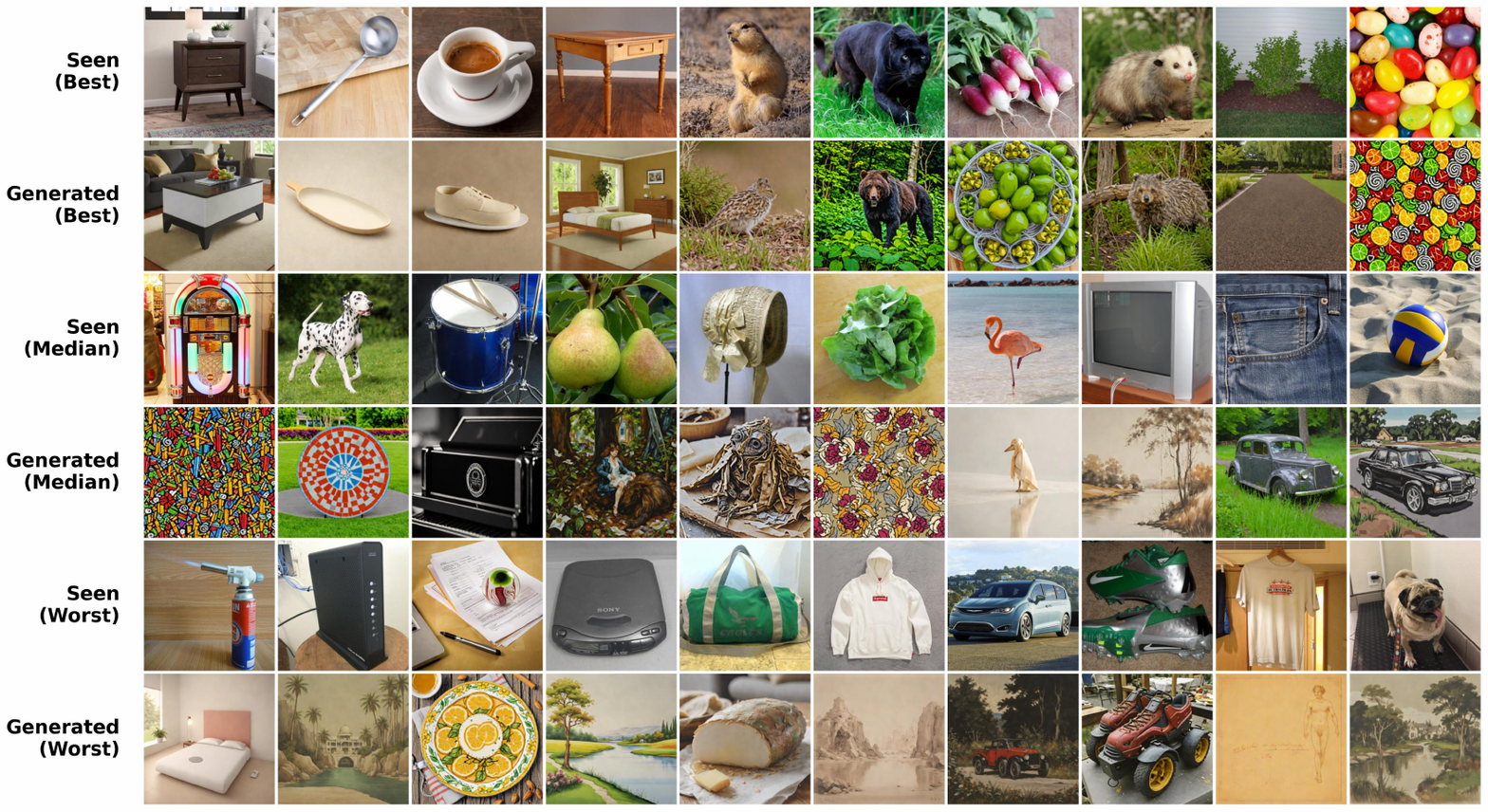}
  \caption{Qualitative results of semantic image reconstruction for \textbf{Subject 09}. The rows correspond to the Best, Median, and Worst reconstruction cases.}
  \label{fig:S11}
\end{figure*}

\begin{figure*}[h!]
  \centering
  \includegraphics[width=0.95\textwidth]{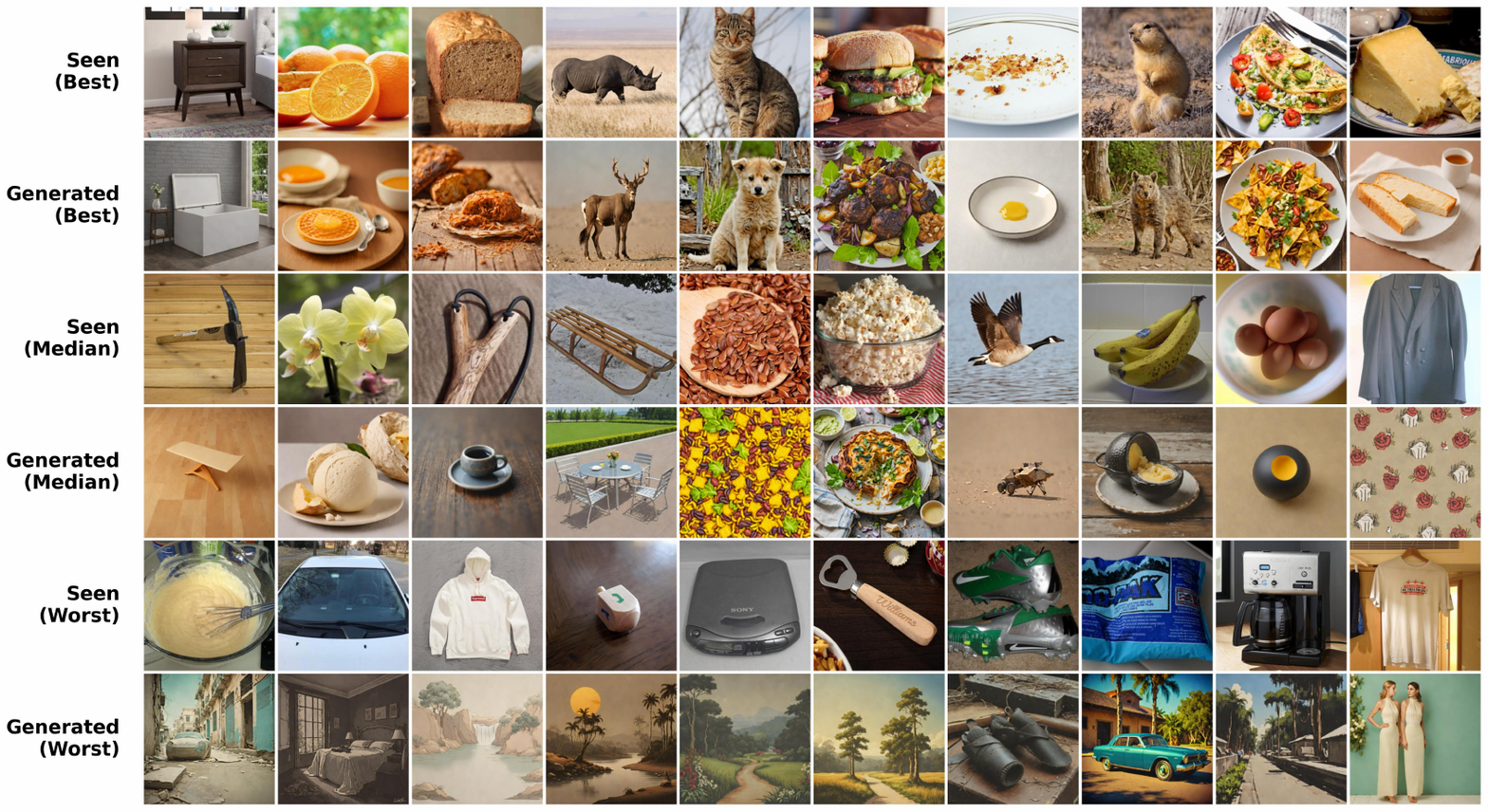}
  \caption{Qualitative results of semantic image reconstruction for \textbf{Subject 10}. The rows correspond to the Best, Median, and Worst reconstruction cases.}
  \label{fig:S12}
\end{figure*}

\clearpage

\begin{figure*}[h!]
  \centering
  \includegraphics[width=0.95\textwidth]{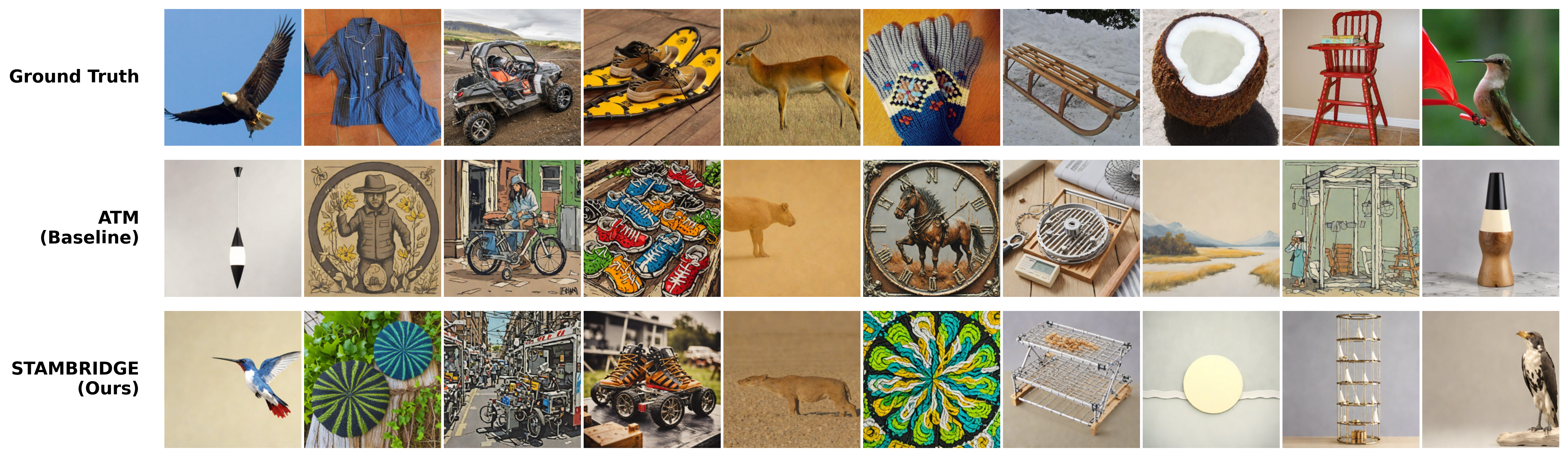}
  \caption{Direct visual comparison between STAMBRIDGE (Ours) and the baseline ATM model for \textbf{Subject 01}. Both models were evaluated using an identical high-level-only generation pipeline. STAMBRIDGE demonstrates a clear advantage in preserving correct overarching semantic categories when the baseline experiences semantic deviation.}
  \label{fig:S13}
\end{figure*}

\begin{figure*}[h!]
  \centering
  \includegraphics[width=0.95\textwidth]{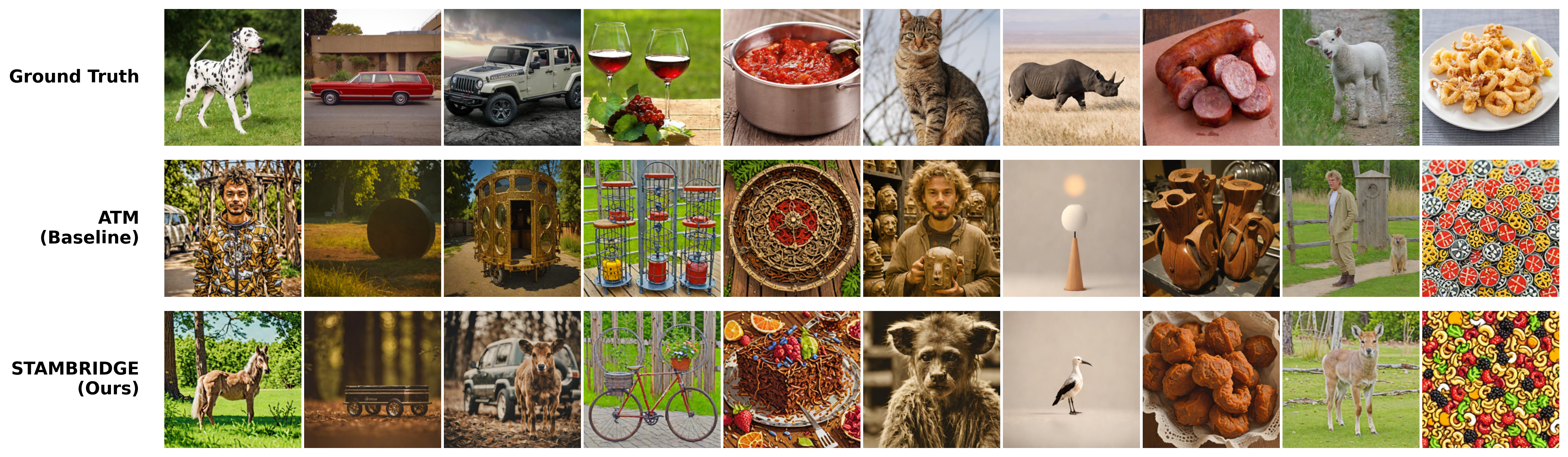}
  \caption{Direct visual comparison between STAMBRIDGE (Ours) and the baseline ATM model for \textbf{Subject 02}. Both models were evaluated using an identical high-level-only generation pipeline. STAMBRIDGE demonstrates a clear advantage in preserving correct overarching semantic categories when the baseline experiences semantic deviation.}
  \label{fig:S14}
\end{figure*}

\begin{figure*}[h!]
  \centering
  \includegraphics[width=0.95\textwidth]{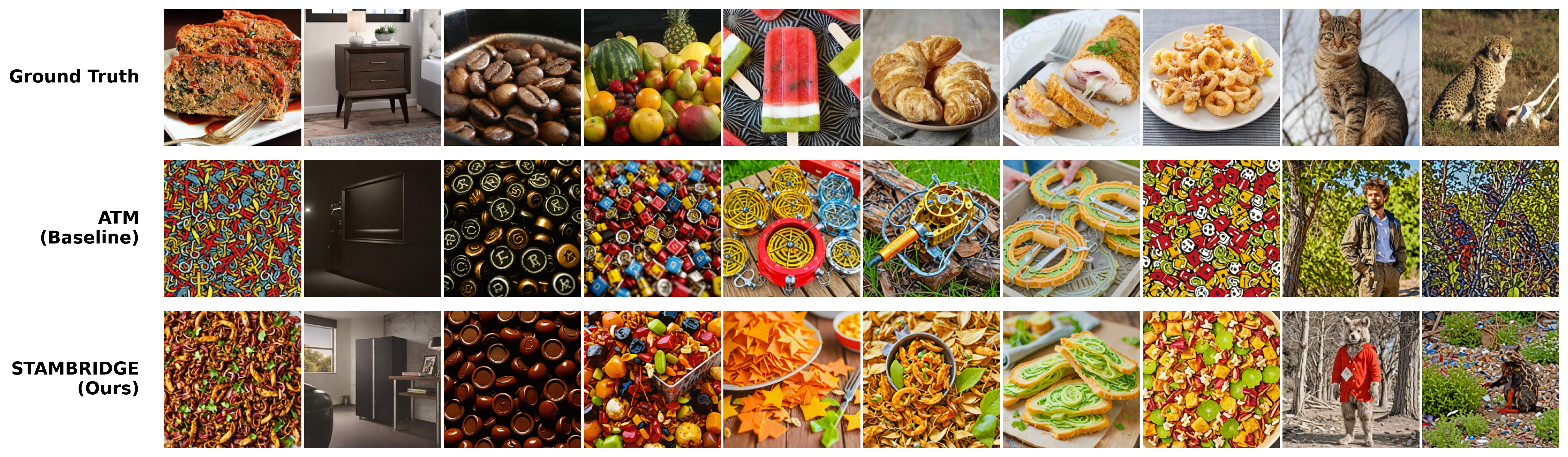}
  \caption{Direct visual comparison between STAMBRIDGE (Ours) and the baseline ATM model for \textbf{Subject 03}. Both models were evaluated using an identical high-level-only generation pipeline. STAMBRIDGE demonstrates a clear advantage in preserving correct overarching semantic categories when the baseline experiences semantic deviation.}
  \label{fig:S15}
\end{figure*}

\begin{figure*}[h!]
  \centering
  \includegraphics[width=0.95\textwidth]{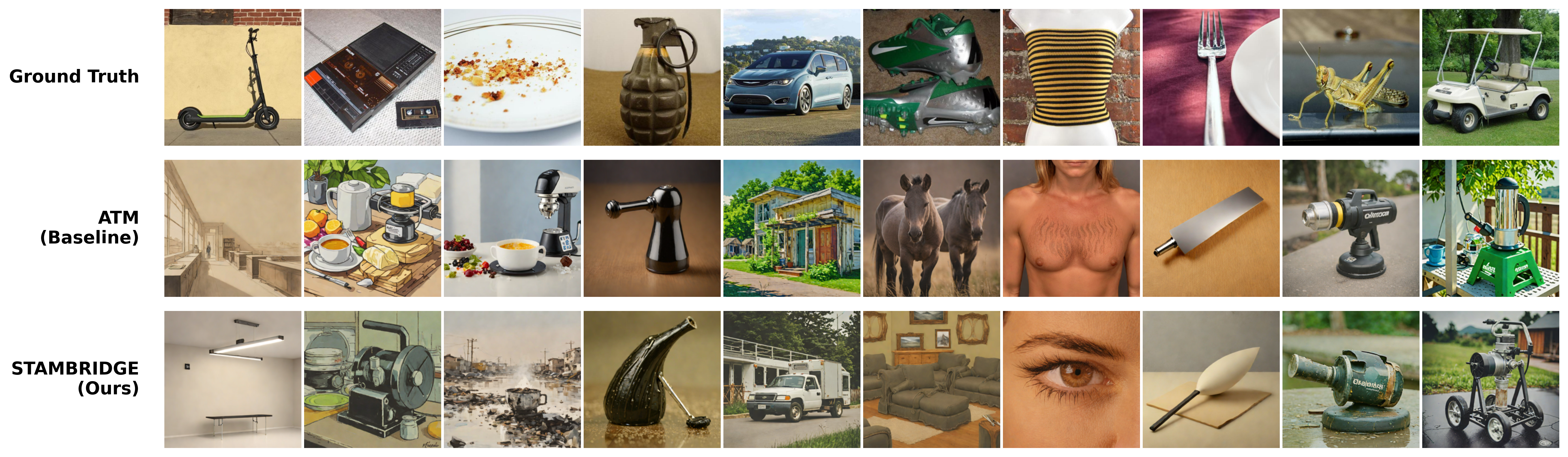}
  \caption{Direct visual comparison between STAMBRIDGE (Ours) and the baseline ATM model for \textbf{Subject 04}. Both models were evaluated using an identical high-level-only generation pipeline. STAMBRIDGE demonstrates a clear advantage in preserving correct overarching semantic categories when the baseline experiences semantic deviation.}
  \label{fig:S16}
\end{figure*}

\begin{figure*}[h!]
  \centering
  \includegraphics[width=0.95\textwidth]{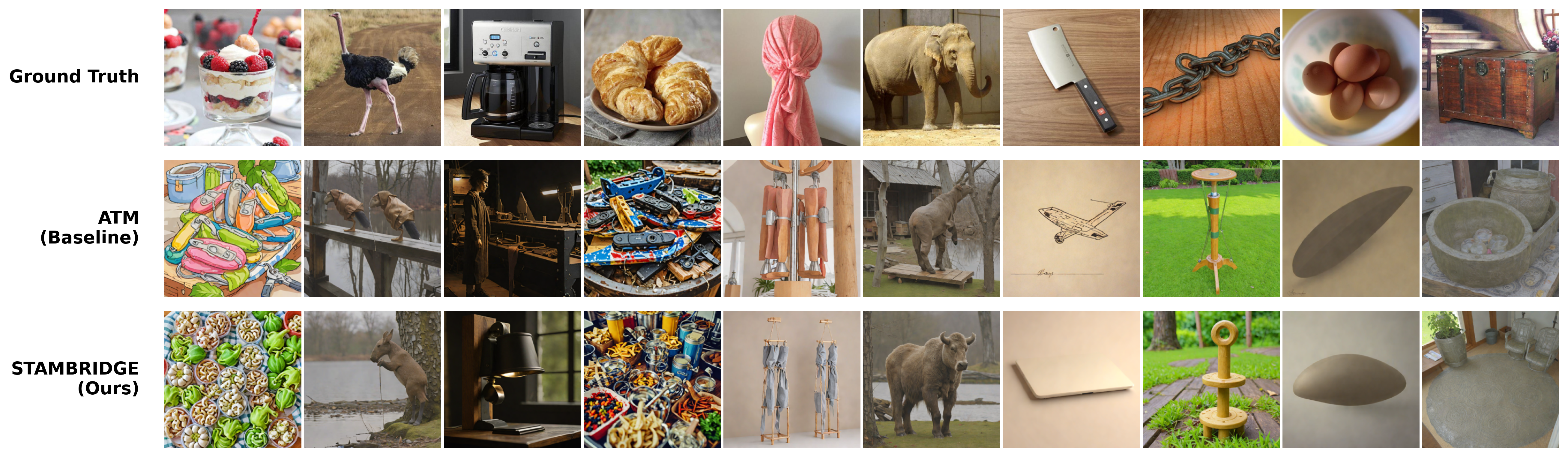}
  \caption{Direct visual comparison between STAMBRIDGE (Ours) and the baseline ATM model for \textbf{Subject 05}. Both models were evaluated using an identical high-level-only generation pipeline. STAMBRIDGE demonstrates a clear advantage in preserving correct overarching semantic categories when the baseline experiences semantic deviation.}
  \label{fig:S17}
\end{figure*}

\begin{figure*}[h!]
  \centering
  \includegraphics[width=0.95\textwidth]{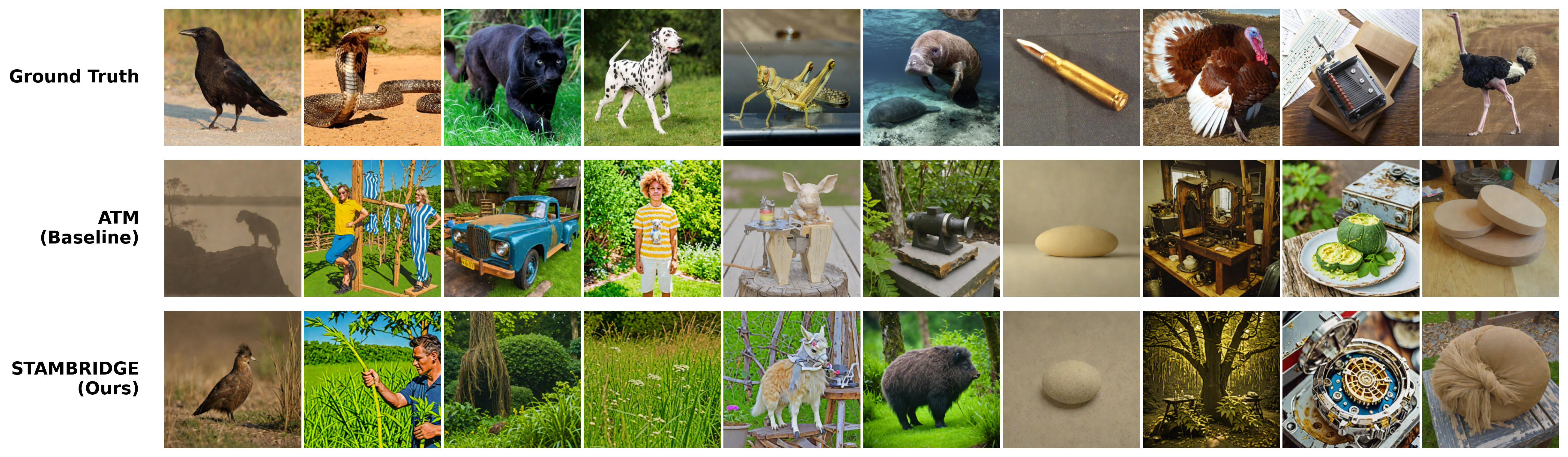}
  \caption{Direct visual comparison between STAMBRIDGE (Ours) and the baseline ATM model for \textbf{Subject 06}. Both models were evaluated using an identical high-level-only generation pipeline. STAMBRIDGE demonstrates a clear advantage in preserving correct overarching semantic categories when the baseline experiences semantic deviation.}
  \label{fig:S18}
\end{figure*}

\begin{figure*}[h!]
  \centering
  \includegraphics[width=0.95\textwidth]{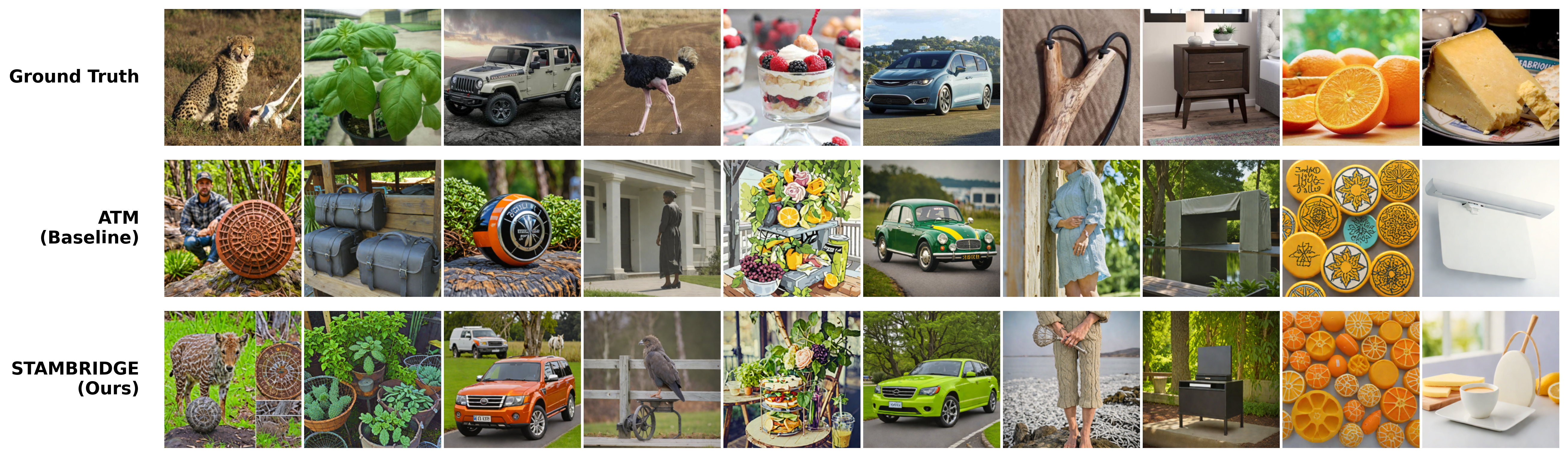}
  \caption{Direct visual comparison between STAMBRIDGE (Ours) and the baseline ATM model for \textbf{Subject 07}. Both models were evaluated using an identical high-level-only generation pipeline. STAMBRIDGE demonstrates a clear advantage in preserving correct overarching semantic categories when the baseline experiences semantic deviation.}
  \label{fig:S19}
\end{figure*}

\begin{figure*}[h!]
  \centering
  \includegraphics[width=0.95\textwidth]{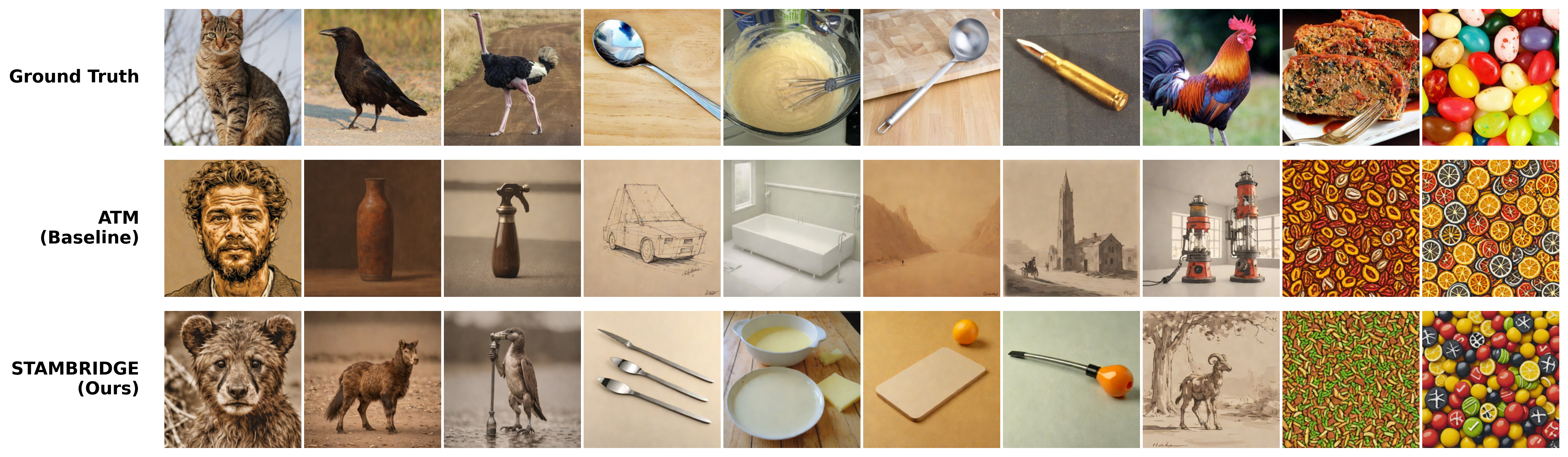}
  \caption{Direct visual comparison between STAMBRIDGE (Ours) and the baseline ATM model for \textbf{Subject 08}. Both models were evaluated using an identical high-level-only generation pipeline. STAMBRIDGE demonstrates a clear advantage in preserving correct overarching semantic categories when the baseline experiences semantic deviation.}
  \label{fig:S20}
\end{figure*}

\begin{figure*}[h!]
  \centering
  \includegraphics[width=0.95\textwidth]{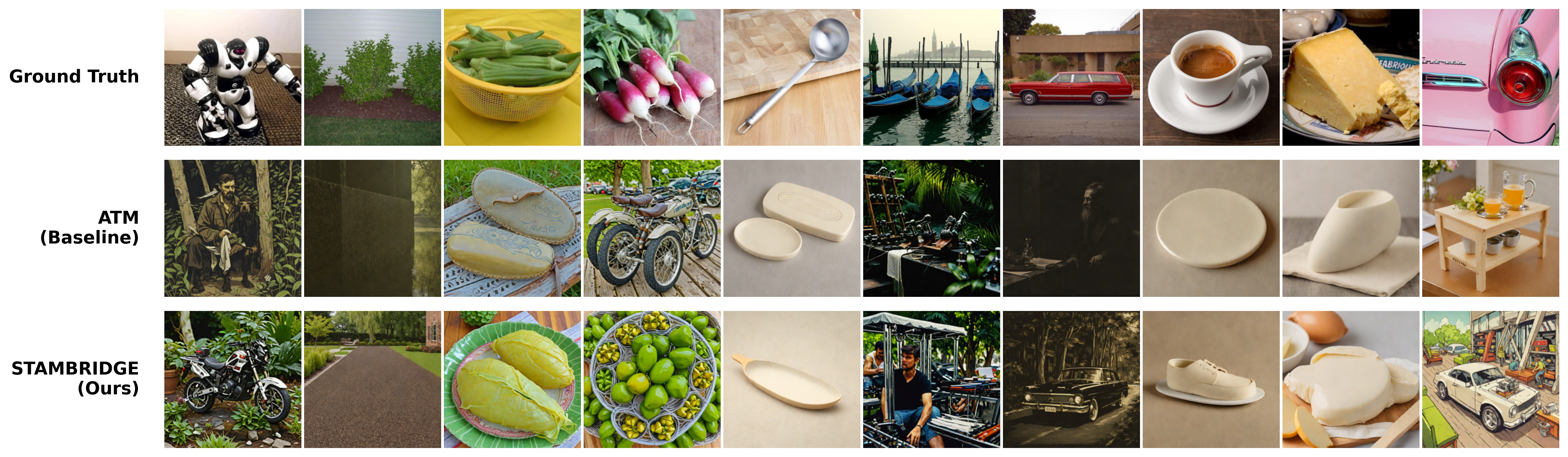}
  \caption{Direct visual comparison between STAMBRIDGE (Ours) and the baseline ATM model for \textbf{Subject 09}. Both models were evaluated using an identical high-level-only generation pipeline. STAMBRIDGE demonstrates a clear advantage in preserving correct overarching semantic categories when the baseline experiences semantic deviation.}
  \label{fig:S21}
\end{figure*}

\begin{figure*}[h!]
  \centering
  \includegraphics[width=0.95\textwidth]{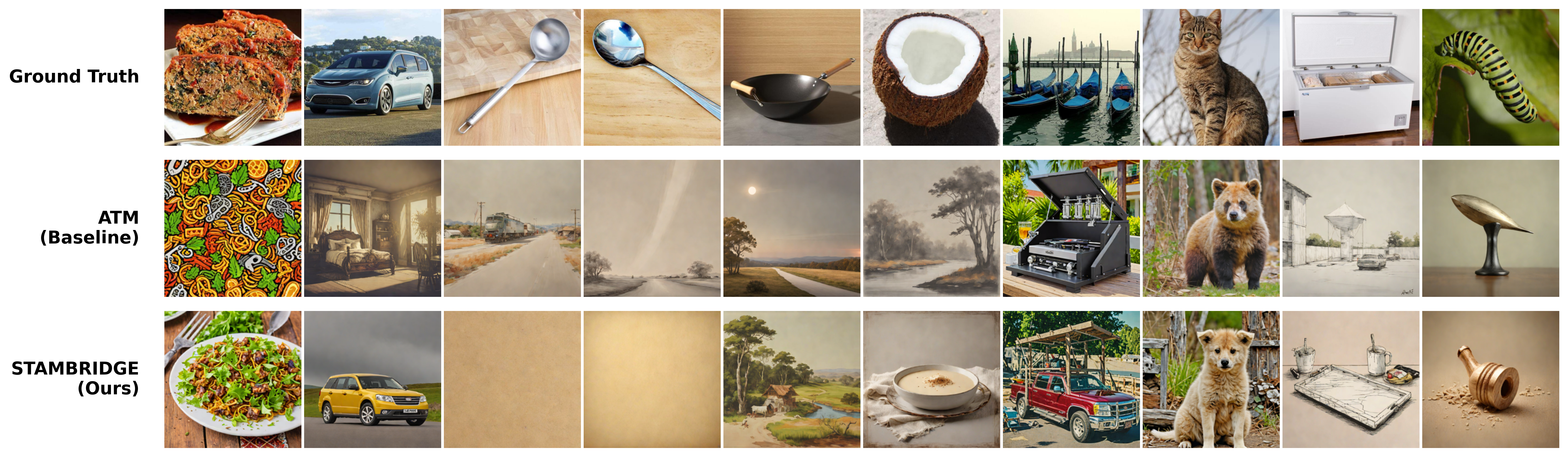}
  \caption{Direct visual comparison between STAMBRIDGE (Ours) and the baseline ATM model for \textbf{Subject 10}. Both models were evaluated using an identical high-level-only generation pipeline. STAMBRIDGE demonstrates a clear advantage in preserving correct overarching semantic categories when the baseline experiences semantic deviation.}
  \label{fig:S22}
\end{figure*}

\clearpage

\begin{figure*}[h!]
  \centering
  \includegraphics[width=\textwidth]{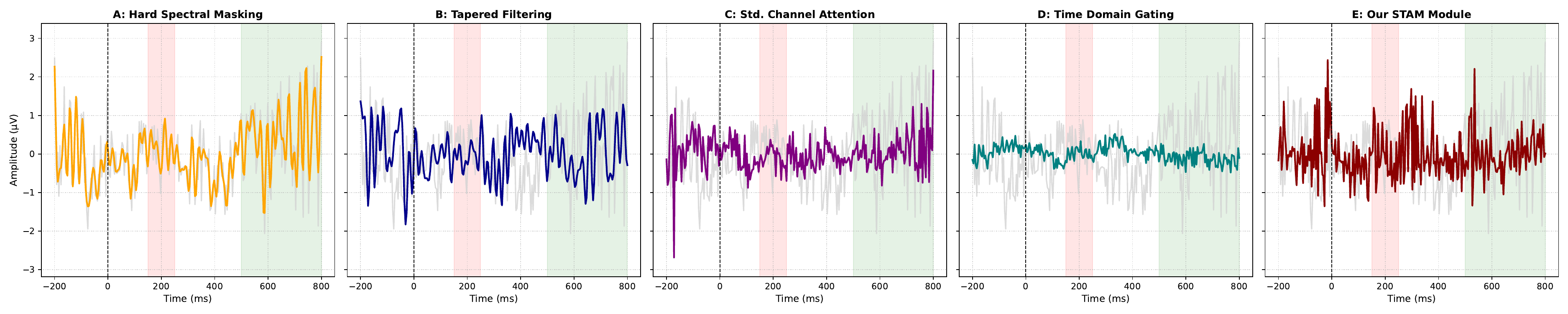}
  \caption{Signal-level comparison of temporal dynamics for \textbf{Subject 01}. Compared to frequency-domain filters (A, B) that introduce ringing, and time/channel gates (C, D) that over-smooth the signal, our STAM module (E) preserves high-frequency transients without artifactual oscillation.}
  \label{fig:wave_sub01}
\end{figure*}

\begin{figure*}[h!]
  \centering
  \includegraphics[width=\textwidth]{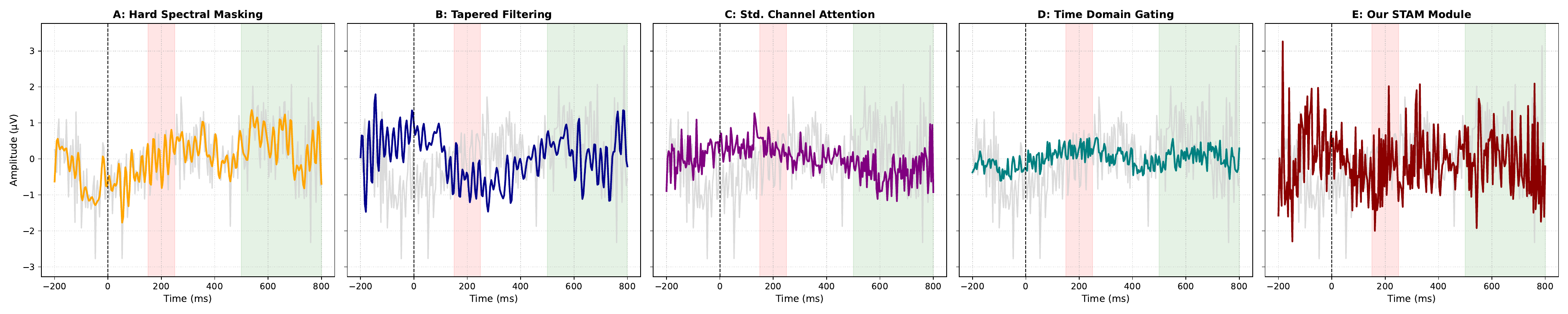}
  \caption{Signal-level comparison of temporal dynamics for \textbf{Subject 02}. The background grey line represents the original raw EEG signal, while the colored lines represent the modulated features. Our STAM module (E) adaptively preserves the original peak transients in the key RSVP time window.}
  \label{fig:wave_sub02}
\end{figure*}

\begin{figure*}[h!]
  \centering
  \includegraphics[width=\textwidth]{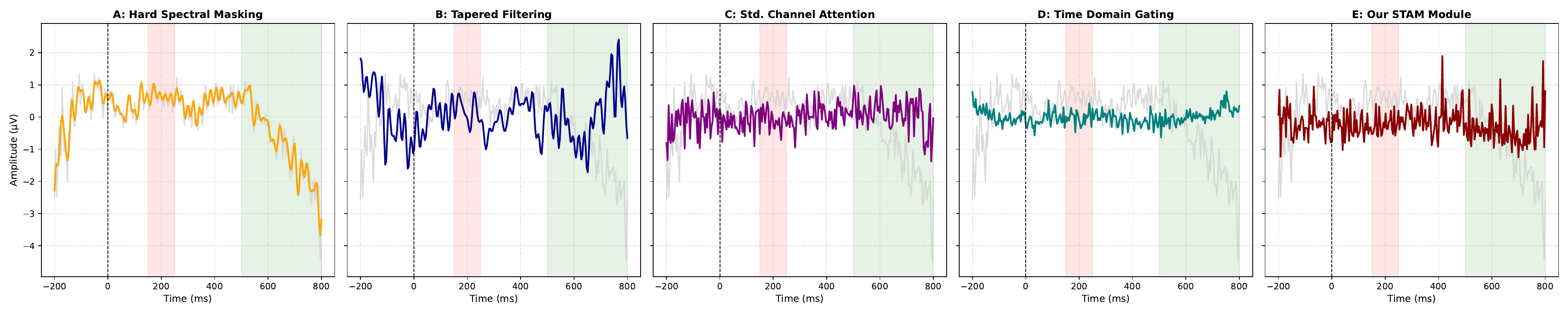}
  \caption{Signal-level comparison of temporal dynamics for \textbf{Subject 03}. Unlike hard spectral masking (A) which induces severe pre-stimulus ripples, STAM (E) maintains a stable baseline.}
  \label{fig:wave_sub03}
\end{figure*}

\begin{figure*}[h!]
  \centering
  \includegraphics[width=\textwidth]{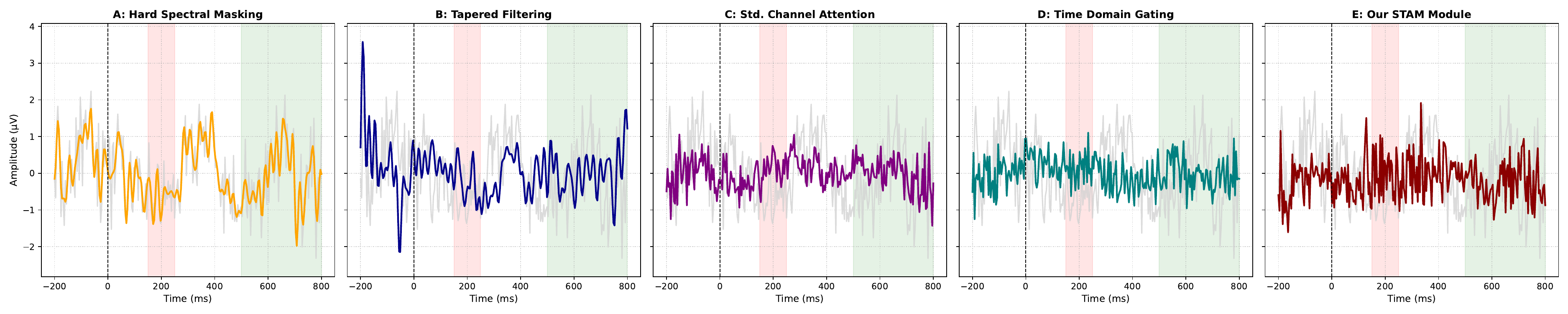}
  \caption{Signal-level comparison of temporal dynamics for \textbf{Subject 04}. STAM (E) dynamically weights the channel features, avoiding the excessive amplitude suppression seen in standard channel attention (C).}
  \label{fig:wave_sub04}
\end{figure*}

\begin{figure*}[h!]
  \centering
  \includegraphics[width=\textwidth]{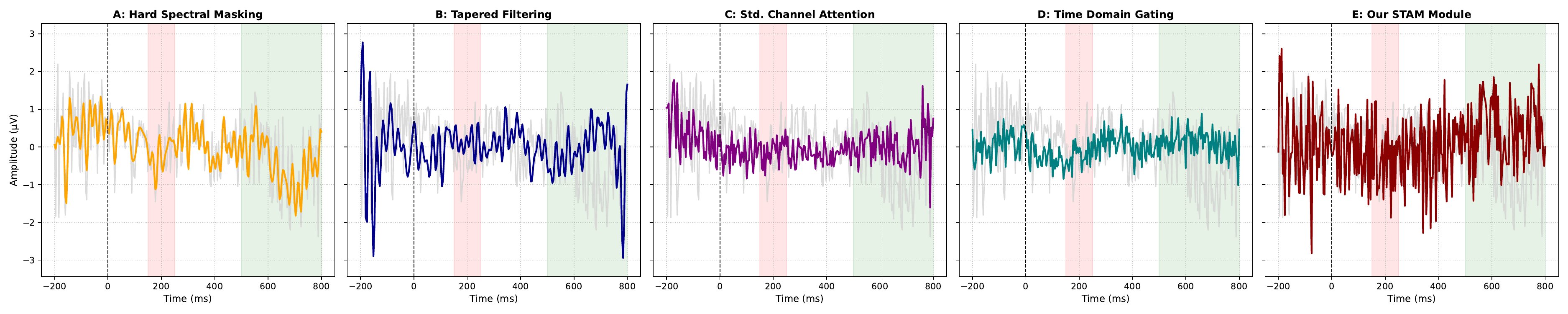}
  \caption{Signal-level comparison of temporal dynamics for \textbf{Subject 05}. The STAM module (E) demonstrates superior capability in capturing sharp neurophysiological transient spikes compared to baselines.}
  \label{fig:wave_sub05}
\end{figure*}

\begin{figure*}[h!]
  \centering
  \includegraphics[width=\textwidth]{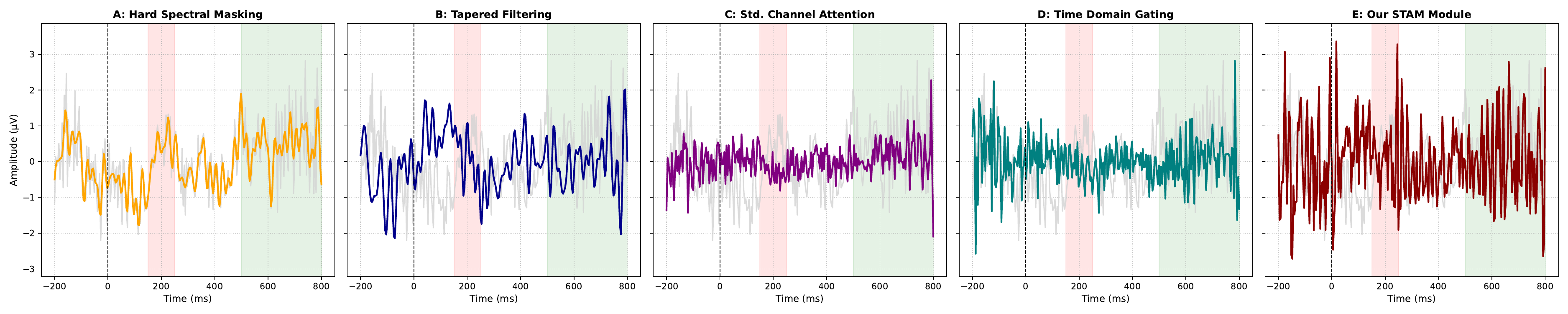}
  \caption{Signal-level comparison of temporal dynamics for \textbf{Subject 06}. Frequency-domain methods (A, B) distort the temporal layout, whereas our amplitude-aware modulation (E) acts smoothly on the time domain.}
  \label{fig:wave_sub06}
\end{figure*}

\begin{figure*}[h!]
  \centering
  \includegraphics[width=\textwidth]{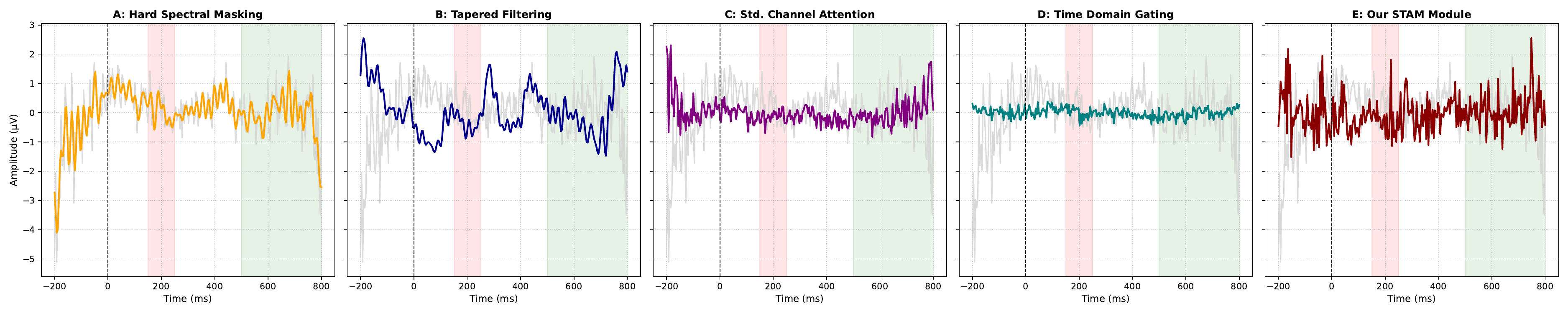}
  \caption{Signal-level comparison of temporal dynamics for \textbf{Subject 07}. STAM (E) successfully preserves the high-frequency components embedded in the evoked potentials.}
  \label{fig:wave_sub07}
\end{figure*}

\begin{figure*}[h!]
  \centering
  \includegraphics[width=\textwidth]{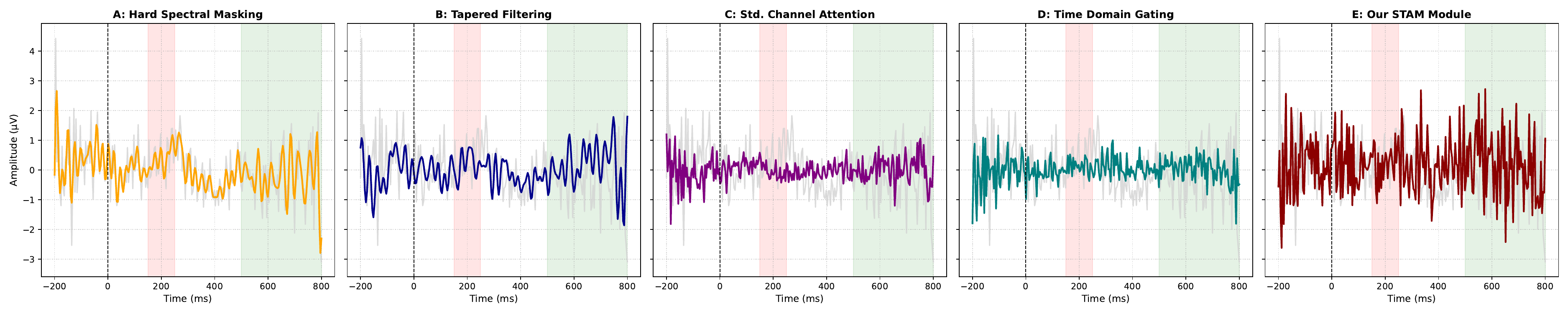}
  \caption{Signal-level comparison of temporal dynamics for \textbf{Subject 08}. The preservation of transients explicitly aids the network in capturing fine-grained semantics during rapid stimulus streams.}
  \label{fig:wave_sub08}
\end{figure*}

\begin{figure*}[h!]
  \centering
  \includegraphics[width=\textwidth]{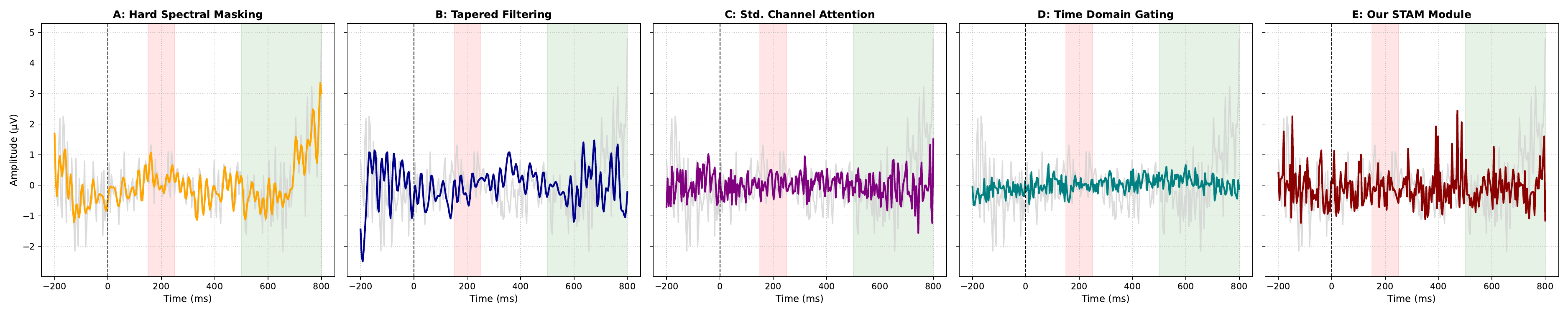}
  \caption{Signal-level comparison of temporal dynamics for \textbf{Subject 09}. STAM (E) maintains high fidelity to the original signal geometry (grey baseline) in key task-relevant intervals.}
  \label{fig:wave_sub09}
\end{figure*}

\begin{figure*}[h!]
  \centering
  \includegraphics[width=\textwidth]{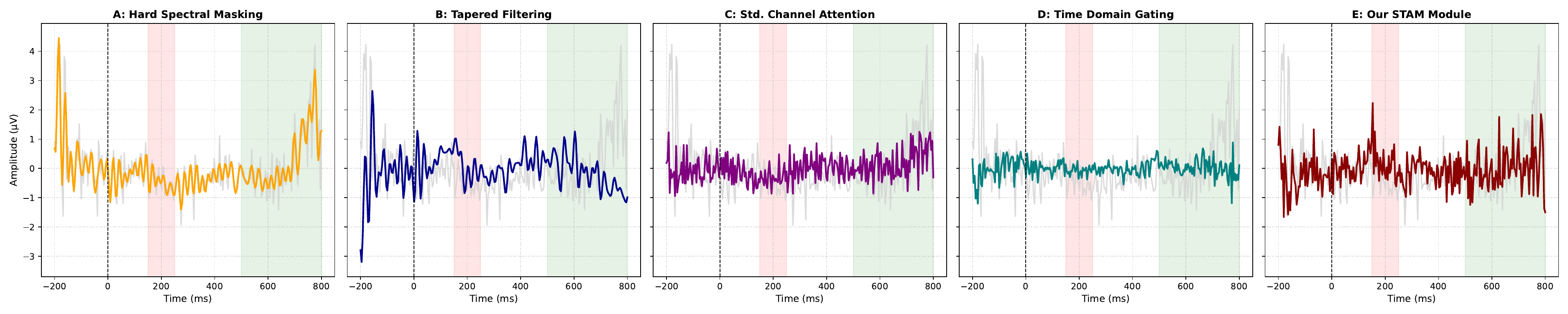}
  \caption{Signal-level comparison of temporal dynamics for \textbf{Subject 10}. Across all subjects, STAM (E) consistently proves to be a non-destructive, feature-preserving modulation strategy.}
  \label{fig:wave_sub10}
\end{figure*}
\end{document}